\documentclass[twocolumn,pra,showpacs]{revtex4}
\usepackage{amssymb}
\usepackage{amsfonts}
\usepackage{amsmath}
\usepackage{graphicx}

\begin{document}

\title{Non-Markovian quantum jumps from measurements in bipartite Markovian
dynamics}
\author{Adri\'{a}n A. Budini$^{1,2}$}
\affiliation{$^{1}$Consejo Nacional de Investigaciones Cient\'{\i}ficas y T\'{e}cnicas
(CONICET), Centro At\'{o}mico Bariloche, Avenida E. Bustillo Km 9.5, (8400)
Bariloche, Argentina}
\affiliation{$^{2}$Universidad Tecnol\'{o}gica Nacional (UTN-FRBA), Fanny Newbery 111,
(8400) Bariloche, Argentina}
\date{\today}

\begin{abstract}
The quantum jump approach allows to characterize the stochastic dynamics
associated to an open quantum system submitted to a continuous measurement
action. In this paper we show that this formalism can consistently be
extended to non-Markovian system dynamics. The results rely in studying a
measurement process performed on a bipartite arrangement characterized by a
Markovian Lindblad evolution. Both a renewal and non-renewal extensions are
found. The general structure of non-local master equations that admit an
unravelling in terms of the corresponding non-Markovian trajectories are
also found. Studying a two-level system dynamics, it is demonstrated that
non-Markovian effects such as an environment-to-system flow of information
may be present in the ensemble dynamics.
\end{abstract}

\maketitle



\section{Introduction}

One of the central achievement of the theory of open quantum systems is the
possibility of assigning to a given master equation an ensemble of
stochastic realizations. They can be put in one-to-one correspondence with a
well defined continuous-in-time measurement process performed over the
system of interest. When the measurement apparatus is sensible to (detect)
transitions between the system's levels \cite%
{zoller,carmichael,dalibard,blatt,heger}, the realizations consist in a
sequence of disruptive instantaneous changes, associated to the measurement
recording events, while in the intermediate time regime the ensemble
dynamics is smooth, being defined by a non-unitary dynamics. These basic
ingredients, which define the quantum jump approach\ (QJA) \cite%
{breuerbook,carmichaelbook,plenio}, are well understood for Markovian
dynamics, that is, those where the evolution of the system density matrix is
local in time.

In the last ten years, an ever increasing interest has been paid to the
development of a consistent non-Markovian generalization of the standard
(Markovian) open quantum system theory \cite{breuerbook}. In the generalized
scheme the system density matrix evolution is characterized by
(time-convoluted) memory contributions \cite%
{stenholm,classBu,Lidar,GrigoBu,VacCol,wilkie,cresser,ManisPetru,vacchini,rate,breuerRate,SemiMarkov,Kosa}%
. Both a theoretical interest as well as a wide range of physical
applications motivates this line of research.

Relevant achievements in the study of non-Markovian master equations were
formulated on the basis of stochastic phenomenological approaches \cite%
{classBu,Lidar,GrigoBu,VacCol} and related concepts \cite%
{wilkie,cresser,ManisPetru,vacchini,rate,breuerRate,SemiMarkov,Kosa}. On the
other hand, much less progress was achieved in the formulation of stochastic
processes that can be read as the result of a continuous measurement action
performed over a system characterized by a non-local in time (non-Markovian)
evolution. In fact, while there exist different stochastic dynamics that in
average recover a non-Markovian density matrix evolution, its reading in
terms of a continuous measurement process is problematic. Remarkable
examples are the non-Markovian quantum state diffusion model \cite{diosi}
and the unravelling of local in time master equations characterized by
negative transition rates \cite{maniscalco}. The realizations associated to
these approaches can only be read in the context of hidden-like variables
models \cite{diosi,maniscalco}.

The main goal of this paper is to demonstrate that it is possible to
formulate a consistent generalization of the QJA such that in average the
ensemble of measurement realizations recover a non-local non-Markovian
density matrix evolution. The basic idea of our analysis is to study the QJA
in a bipartite Markovian arrange. Then, we search for the conditions
(interaction symmetries) that allows to formulate a closed stochastic
dynamic for the system of interest. The coupling with the second or
auxiliary system introduce the memory effects. In contrast with previous
approaches \cite{diosi,maniscalco}, the reading of the stochastic
realizations in terms of a continuous-in-time measurement process is
guaranteed by construction.

We show that a renewal non-Markovian measurement process can be obtained
from the bipartite dynamics. Renewal means that the 
interval statistics between successive events is always the same being
defined by a probability distribution called waiting time distribution \cite%
{carmichael}. A non-renewal dynamics is also defined. As in the standard
Markovian formalism, the arising of each case depends on the properties of
the resetting state \cite{heger} associated to each measurement event. The
structure of the corresponding non-Markovian master equations are also found.

We remark that there exist previous studies where the QJA is formulated for
a system that interacts with extra unobserved \textquotedblleft
classical\textquotedblright\ degrees of freedom \cite%
{BuJumpSMS,petruccione,barchielli}. While our approach relies on a similar
underlying dynamic (strictly, here not any classicality condition is
imposed), we demonstrate that over a similar basis it is possible to get a
consistent non-Markovian generalization of the QJA. In fact, in contrast
with previous contributions \cite{BuJumpSMS,petruccione,barchielli}, we
focus the analysis on the possibility of establishing a closed stochastic
system dynamics, that is, without involving \textquotedblleft
explicitly\textquotedblright\ the degrees of freedom of the auxiliary system.

The paper is outlined as follows. In Sec. II, in order to introduce the
notation as well as basic results over which our analysis rely, we provide a
resume of the standard\ Markovian QJA. In Sec. III we demonstrate that\ the
basic structure of the standard QJA can be embedded in a bipartite Markovian
dynamics, providing in this way the theoretical background for its
non-Markovian generalization. Possible (bipartite) interactions that lead to
a closed system dynamics are found. The non-Markovian density matrix
evolution is determined for both renewal and non-renewal measurement
processes. In Sec. IV we study a particular example that explicitly shows
the consistence of the present proposal. Furthermore, it demonstrate that
non-Markovian features such as an environment-to-system flow of information 
\cite{NoMeasure} may be present in the ensemble dynamics. The conclusions
are presented in Sec. V. In Appendix A we provide a derivation of the
statistics of the measurement events in the standard case. In Appendix B we
work out an alternative derivation of the non-Markovian system density
matrix evolution based on the measurement statistics.

\section{Markovian quantum jumps}

The standard QJA allows to define the (stochastic) dynamics of an open
quantum system when it is subjected to a measurement process. The basic
ingredients of the formalism are the system density matrix evolution, the
definition of the apparatus measurement action, the conditional dynamic
between detections events and their statistical characterization. Below, we
review these elements.

We write the evolution of the system density matrix $\rho _{t}^{s}$ as%
\begin{equation}
\frac{d}{dt}\rho _{t}^{s}=(\mathcal{\hat{L}}_{0}+\sum_{\alpha }\gamma
_{\alpha }\mathcal{\hat{C}}[V_{\alpha }])\rho _{t}^{s},  \label{MasterMarkov}
\end{equation}%
where $\mathcal{\hat{L}}_{0}$ is an arbitrary superoperator that may include
Hamiltonian as well as dissipative (Lindblad) superoperators \cite%
{breuerbook}. From now on the hat symbol denotes superoperators. The second
contribution in (\ref{MasterMarkov}) is defined by an addition of Lindblad
channels%
\begin{equation}
\mathcal{\hat{C}}[V]\rho =V\rho V^{\dag }-\frac{1}{2}\{V^{\dag }V,\rho
\}_{+},  \label{LindbladDEF}
\end{equation}%
each one characterized by the operator $V_{\alpha }$ and the transition rate 
$\gamma _{\alpha }.$ With $\{\cdot ,\cdot \}_{+}$ we denotes an
anticommutation operation.

We assume that the system is monitored by only one measurement apparatus,
which is sensible to all Lindblad transitions channels $\mathcal{\hat{C}}%
[V_{\alpha }].$ Hence, the master equation (\ref{MasterMarkov}) is rewritten
as%
\begin{equation}
\frac{d}{dt}\rho _{t}^{s}=(\mathcal{\hat{D}}+\mathcal{\hat{J}})\rho _{t}^{s}.
\label{LindbladSA}
\end{equation}%
The superoperator $\mathcal{\hat{J}}$ reads%
\begin{equation}
\mathcal{\hat{J}}\rho =\sum_{\alpha }\gamma _{\alpha }V_{\alpha }\rho
V_{\alpha }^{\dag }.  \label{JMarkov}
\end{equation}%
It defines the system transformation after a measurement event. In fact,
when a recording event happens, consistently with a quantum measurement
theory \cite{breuerbook}, the system density matrix suffer the disruptive
transformation $\rho \rightarrow \mathcal{\hat{M}}\rho $ (jump or state
collapse),%
\begin{equation}
\mathcal{\hat{M}}\rho =\frac{\mathcal{\hat{J}}\rho }{\mathrm{Tr}_{s}[%
\mathcal{\hat{J}}\rho ]}=\frac{\sum_{\alpha }\gamma _{\alpha }V_{\alpha
}\rho V_{\alpha }^{\dag }}{\left\{ \sum_{\alpha }\gamma _{\alpha }\mathrm{Tr}%
_{s}[V_{\alpha }^{\dag }V_{\alpha }\rho ]\right\} },  \label{MOperator}
\end{equation}%
where $\mathrm{Tr}_{s}[\mathcal{\cdots }]$ denotes a trace operation. On the
other hand, in Eq. (\ref{LindbladSA}) the superoperator $\mathcal{\hat{D}}$
is defined as%
\begin{equation}
\mathcal{\hat{D}}\rho =\mathcal{\hat{L}}_{0}\rho -\frac{1}{2}\sum_{\alpha
}\gamma _{\alpha }\{V_{\alpha }^{\dag }V_{\alpha },\rho \}_{+}.
\label{DMarkov}
\end{equation}%
In the QJA, this superoperator defines the system dynamics between detection
events. In fact, given that in the interval $(\tau ,t)$\ not any detection
event happens, the system dynamic is defined by the (conditional) normalized
propagator%
\begin{equation}
\mathcal{\hat{T}}_{c}(t-\tau )\rho =\frac{\mathcal{\hat{T}}(t-\tau )\rho }{%
\mathrm{Tr}_{s}[\mathcal{\hat{T}}(t-\tau )\rho ]}.  \label{Conditional}
\end{equation}%
The superoperator $\mathcal{\hat{D}}$ generates the dynamics of the
unnorrmalized propagator $\mathcal{\hat{T}}(t-\tau ),$ which reads%
\begin{equation}
\mathcal{\hat{T}}(t-\tau )\rho =\exp [(t-\tau )\mathcal{\hat{D}}]\rho .
\label{Texponencial}
\end{equation}%
In this way, the trajectories associated to the measurement process are a
piecewise deterministic process \cite{breuerbook}\ which combine a
deterministic time-evolution [Eq. (\ref{Conditional})] with jump process
[Eq. (\ref{MOperator})].

The propagator $\mathcal{\hat{T}}(t)$ completely define the statistics of
the measurement process. In fact, it allows to calculate the survival
probability between measurement events. Given that at time $\tau $ the state
of the system is $\rho _{\tau },$ the probability $P_{0}(t-\tau |\rho _{\tau
})$ of not happening any detection event in the interval $(\tau ,t)$\ is%
\begin{equation}
P_{0}(t-\tau |\rho _{\tau })=\mathrm{Tr}_{s}[\mathcal{\hat{T}}(t-\tau )\rho
_{\tau }].  \label{SurMarkov}
\end{equation}%
The probability distribution $w(t-\tau |\rho _{\tau })$ of the interval $%
(t-\tau )$ follows as $w(t-\tau |\rho _{\tau })=-(d/dt)P_{0}(t-\tau |\rho
_{\tau }),$ delivering%
\begin{equation}
w(t-\tau |\rho _{\tau })=-\mathrm{Tr}_{s}[\mathcal{\hat{D}\hat{T}}(t-\tau
)\rho _{\tau }].  \label{WaitMarkov}
\end{equation}%
By using that $(d/dt)\mathrm{Tr}_{s}[\rho _{t}^{s}]=0,$ Eq. (\ref{LindbladSA}%
) implies that $-\mathrm{Tr}_{s}[\mathcal{\hat{D}}\cdot ]=\mathrm{Tr}_{s}[%
\mathcal{\hat{J}}\cdot ],$ leading to the equivalent expression $w(t-\tau
|\rho _{\tau })=\mathrm{Tr}_{s}[\mathcal{\hat{J}\hat{T}}(t-\tau )\rho _{\tau
}].$ On the other hand, notice that $P_{0}(t-\tau |\rho _{\tau }),$ or
equivalently $w(t-\tau |\rho _{\tau }),$ depends explicitly on the state $%
\rho _{\tau }.$

From the previous statistical objects it is possible to define the
\textquotedblleft conditional distribution\textquotedblright\ \cite%
{carmichael}%
\begin{equation}
w_{c}(t-\tau |\rho _{\tau })=\frac{w(t-\tau |\rho _{\tau })}{P_{0}(t-\tau
|\rho _{\tau })}.  \label{WaitConditional}
\end{equation}%
It defines the probability density for recording a detection event at time $%
t,$ given no counts are recorded in the interval $(\tau ,t),$ and given that
the last one was recorded at time $\tau .$ Therefore, $w_{c}(t-\tau |\rho
_{\tau })$ gives the probability density for a jump at time $t$ given that
we know that no event occurred up to the present time since the last one 
\cite{carmichael}. Trivially, from Eqs. (\ref{SurMarkov}) and (\ref%
{WaitMarkov}) it can be written as%
\begin{equation}
w_{c}(t-\tau |\rho _{\tau })=\frac{-\mathrm{Tr}_{s}[\mathcal{\hat{D}\hat{T}}%
(t-\tau )\rho _{\tau }]}{\mathrm{Tr}_{s}[\mathcal{\hat{T}}(t-\tau )\rho
_{\tau }]}.  \label{WaitConditionalMarkov}
\end{equation}

\subsection{Stochastic dynamics}

With the previous elements, it is possible to define the dynamics of a
stochastic density matrix $\rho _{s}^{\mathrm{st}}(t)$ such that its average
over realizations, denoted by an overbar, recovers the system state%
\begin{equation}
\rho _{t}^{s}=\overline{\rho _{s}^{\mathrm{st}}(t)}.  \label{SPromedio}
\end{equation}%
Each realization corresponds to a given recording realization of the
measurement apparatus. Its structure can be established by studying the
counting statistics of the measurement process (see Appendix A).

Given the initial state $\rho _{0}^{s},$ we can evaluate $P_{0}(t-0|\rho
_{0}^{s}).$ The time $t_{1}$ of the first detection event follows by solving
the equation $P_{0}(t_{1}-0|\rho _{0}^{s})=r,$ where $r$ is a random number
in the interval $(0,1).$ The dynamic of $\rho _{s}^{\mathrm{st}}(t)$\ in the
interval $(0,t_{1})$ is defined by Eq. (\ref{Conditional}). At $t=t_{1}$ the
disruptive transformation $\rho _{s}^{\mathrm{st}}(t_{1})\rightarrow 
\mathcal{\hat{M}}\rho _{s}^{\mathrm{st}}(t_{1})$ is applied. The subsequent
dynamics is the same. In fact, after the $n_{th}-$measurement event at time $%
t_{n},$ $\rho _{s}^{\mathrm{st}}(t_{n})\rightarrow \mathcal{\hat{M}}\rho
_{s}^{\mathrm{st}}(t_{n}),$ the time $t_{n+1}$ for the next detection event
follows from%
\begin{equation}
P_{0}(t_{n+1}-t_{n}|\mathcal{\hat{M}}\rho _{s}^{\mathrm{st}}(t_{n})),
\label{PEstocastica}
\end{equation}%
equated to $r,$ where again $r$ is a random number in the interval $(0,1).$
The dynamic in the interval $(t_{n},t_{n+1})$ is defined by the conditional
propagator (\ref{Conditional}).

The previous algorithm determine the realizations over finite time intervals 
\cite{plenio}. It is also possible to obtain the evolution over
infinitesimal intervals. Its structure remains the same [Eqs. (\ref%
{MOperator}) and (\ref{Conditional})]. Nevertheless, instead of Eq. (\ref%
{PEstocastica}), the jump statistic is determined from $w_{c}(t-\tau |\rho
_{\tau }),$ Eq. (\ref{WaitConditional}). Given that the last event happened
at time $\tau $ and that not any detection was detected in the interval $%
(\tau ,t),$ the probability $\Delta P$ of having a detection event in the 
\textit{infinitesimal interval} $(t,t+dt)$ is (by definition) \cite%
{carmichael}%
\begin{equation}
\Delta P=w^{c}(t-\tau |\rho _{s}^{\mathrm{st}}(\tau ))\ dt.
\label{DeltaPDefinicion}
\end{equation}%
From Eqs. (\ref{DMarkov}), (\ref{Conditional}), and (\ref%
{WaitConditionalMarkov}), we can write%
\begin{equation}
\Delta P\!=\!-dt\mathrm{Tr}_{s}[\mathcal{\hat{D}}\rho _{s}^{\mathrm{st}%
}(t)]\!=\!dt\sum_{\alpha }\gamma _{\alpha }\mathrm{Tr}_{s}[V_{\alpha }^{\dag
}V_{\alpha }\rho _{s}^{\mathrm{st}}(t)].  \label{DeltaPMarkov}
\end{equation}%
The happening or not of a detection follows by comparing $\Delta P$ with a
random number in the interval $(0,1).$ This alternative algorithm generate
the same realizations than the previous one \cite{plenio}. Nevertheless, in
this last scheme the Markovian property of the underlying master equation is
self-evident in the expression of $\Delta P.$ In fact, $\Delta P$ does not
depends on the \textquotedblleft history\textquotedblright\ of $\rho _{s}^{%
\mathrm{st}}(t)$ in the interval $(\tau ,t).$ It only depends on $\rho _{s}^{%
\mathrm{st}}(t).$

\subsection{Renewal and non-renewal measurement processes}

An extra understanding of the QJA is achieved by specifying the operators $%
\{V_{\alpha }\}$ that determine the measurement transformation Eq. (\ref%
{MOperator}).

When the system state after a measurement event (resetting state) is always
the same, the statistics of the time interval between events is defined by a
unique probability distribution (waiting time distribution). In this case,
the measurement process is a renewal one. This situation arises when the
measurement apparatus is sensible to all transitions $(\left\vert
u\right\rangle \rightsquigarrow \left\vert r_{\alpha }\right\rangle )$\
between a given system state $\left\vert u\right\rangle $\ and a set of
alternative states $\{\left\vert r_{\alpha }\right\rangle \}.$ Therefore,
the operators $\{V_{\alpha }\}$ have the structure%
\begin{equation}
V_{\alpha }=\left\vert r_{\alpha }\right\rangle \left\langle u\right\vert ,
\label{ValfaRenewal}
\end{equation}%
which in turn, from Eq. (\ref{MOperator}), imply the measurement
transformation%
\begin{equation}
\mathcal{\hat{M}}\rho =\bar{\rho}_{s}\equiv \sum_{\alpha }p_{\alpha
}\left\vert r_{\alpha }\right\rangle \left\langle r_{\alpha }\right\vert ,\
\ \ \ \ \ \ p_{\alpha }=\frac{\gamma _{\alpha }}{\{\sum_{\alpha }\gamma
_{\alpha }\}}.  \label{Mrenewal}
\end{equation}%
Hence, the conditional dynamics [Eq. (\ref{Conditional})] always start in
the same resetting state $\bar{\rho}_{s}$\ \cite{heger,plenio}.\
Furthermore, the survival probability and waiting time distribution [Eqs. (%
\ref{SurMarkov}) and (\ref{WaitMarkov}) respectively], after the first event 
$[\rho _{\tau }\rightarrow \mathcal{\hat{M}}\rho =\bar{\rho}_{s}]$ are
always the same, being defined as%
\begin{equation}
P_{0}(t)=\mathrm{Tr}_{s}[\mathcal{\hat{T}}(t)\bar{\rho}_{s}],\ \ \ \ \ \ \ \
\ \ \ \ w(t)=-\mathrm{Tr}_{s}[\mathcal{\hat{D}\hat{T}}(t)\bar{\rho}_{s}].
\label{MarkovRenewal}
\end{equation}%
In consequence, the interval statistics does not depends explicitly on the
time $\tau $\ of the last events and it is always the same. The operators (%
\ref{ValfaRenewal}) arise for example in optical systems such as two-level
fluorescent systems, where $\bar{\rho}_{s}$ is a pure state, and
three-level\ $\Lambda $ configurations \cite{heger,plenio}.

In general, the operators may read%
\begin{equation}
V_{\alpha }=\left\vert r_{\alpha }\right\rangle \left\langle u_{\alpha
}\right\vert ,  \label{ValfaNoRenewal}
\end{equation}%
that is, the measurement apparatus is sensible to different transitions $%
\left\vert u_{\alpha }\right\rangle \rightsquigarrow \left\vert r_{\alpha
}\right\rangle .$ This case may happen when the natural frequencies of the
different transitions are indistinguishable for the measurement apparatus;
for example in cascade optical systems \cite{heger}. The measurement
transformation%
\begin{equation}
\mathcal{\hat{M}}\rho =\frac{\sum_{\alpha }\gamma _{\alpha }\left\langle
u_{\alpha }\right\vert \rho \left\vert u_{\alpha }\right\rangle \left\vert
r_{\alpha }\right\rangle \left\langle r_{\alpha }\right\vert }{\left\{
\sum_{\alpha }\gamma _{\alpha }\left\langle u_{\alpha }\right\vert \rho
\left\vert u_{\alpha }\right\rangle \right\} },  \label{MNoRenewal}
\end{equation}%
delivers a state that depends on the pre-detection state. Hence, it is not
possible to define a unique statistical object as in the previous case, that
is, the survival probability and waiting time distribution correspond to the
general expressions Eqs. (\ref{SurMarkov}) and (\ref{WaitMarkov})
respectively.

\section{Non-Markovian quantum jumps from bipartite Markovian dynamics}

The previous elements and results that define the QJA, without introducing
any new element, can also be\ established for bipartite dynamics. Here, in
addition to the system of interest $S$ we consider an auxiliary or ancilla
system $A.$ Their joint dynamics is Markovian. Furthermore, we assume that
the measurement apparatus is sensible to the same system transitions as
before. Thus, we can define a stochastic density matrix $\rho _{\mathrm{st}%
}^{sa}(t)$ such that its average over realizations recovers the bipartite
density matrix $\rho _{t}^{sa}=\overline{\rho _{\mathrm{st}}^{sa}(t)}.$ The
density matrix of $S$ is recovered by a partial trace operation over the
auxiliary system $A,$%
\begin{equation}
\rho _{t}^{s}=\mathrm{Tr}_{a}[\rho _{t}^{sa}]=\mathrm{Tr}_{a}[\overline{\rho
_{\mathrm{st}}^{sa}(t)}].  \label{RhoPartialSistema}
\end{equation}%
Trivially, by introducing the stochastic matrix%
\begin{equation}
\rho _{\mathrm{st}}^{s}(t)=\mathrm{Tr}_{a}[\rho _{\mathrm{st}}^{sa}(t)],
\label{RhoSystemEstocastica}
\end{equation}%
we recover Eq. (\ref{SPromedio}), that is, $\rho _{t}^{s}=\overline{\rho _{%
\mathrm{st}}^{s}(t)}.$ At this point, we ask about the existence of
different $S-A$ interactions an evolutions under which it is possible to get
a closed stochastic dynamics for $\rho _{\mathrm{st}}^{s}(t),$ that is,\
without involving explicitly the ancilla state. In addition to this
constraint, here we search interaction structures that introduce a minimal
modification of the standard approach, that is, it should be possible to
define a measurement transformation [Eq. (\ref{MOperator})], a conditional
interevent dynamic [Eq. (\ref{Conditional})], and a survival probability
[Eq. (\ref{SurMarkov})].

\subsection{Bipartite Markovian embedding}

Taking into account the evolution Eq. (\ref{MasterMarkov}), we write the
bipartite evolution as%
\begin{equation}
\frac{d}{dt}\rho _{t}^{sa}=(\mathbb{\hat{L}}_{0}+\sum_{\alpha lm}\gamma
_{\alpha lm}\mathcal{\hat{C}}[V_{\alpha lm}])\rho _{t}^{sa}.
\label{LindbladBipartito}
\end{equation}%
The operators $V_{\alpha lm}$ are defined as%
\begin{equation}
V_{\alpha lm}=V_{\alpha }\otimes \left\vert a_{l}\right\rangle \left\langle
a_{m}\right\vert .  \label{VAlfaLM}
\end{equation}%
The set of states $\{\left\vert a_{l}\right\rangle \}$ provides an
orthogonal and normalized basis of the ancilla Hilbert space. The system
operators $\{V_{\alpha }\}$ are the same as before. Notice that the diagonal
contributions, defined by the operators $V_{\alpha mm}=V_{\alpha }\otimes
\left\vert a_{m}\right\rangle \left\langle a_{m}\right\vert ,$ correspond to
system's transitions that only happen when the ancilla system is in the
state $\left\vert a_{m}\right\rangle .$ The non-diagonal contributions $%
V_{\alpha lm}=V_{\alpha }\otimes \left\vert a_{l}\right\rangle \left\langle
a_{m}\right\vert $ correspond to system transitions that occur
simultaneously with the ancilla transition $\left\vert a_{m}\right\rangle
\rightsquigarrow \left\vert a_{l}\right\rangle .$

In Eq. (\ref{LindbladBipartito}), the superoperator $\mathbb{\hat{L}}_{0}$
not only includes the system evolution [$\mathcal{\hat{L}}_{0}$ in Eq. (\ref%
{MasterMarkov})] but also an arbitrary evolution for the ancilla system as
well as the system-ancilla interaction. Over this last contribution, we only
demand that it must not to include any interaction proportional to the
transitions defined by the operators $\{V_{\alpha }\}.$ On the other hand, \
the measurement apparatus remains the same, that is, it only detects the
system transitions. Therefore, we split the bipartite master equation (\ref%
{LindbladBipartito}) as%
\begin{equation}
\frac{d}{dt}\rho _{t}^{sa}=(\mathbb{\hat{D}}+\mathbb{\hat{J}})\rho _{t}^{sa},
\label{BipartitaD+JMarkkov}
\end{equation}%
where the superoperator $\mathbb{\hat{J}}$ reads%
\begin{equation}
\mathbb{\hat{J}}\rho =\sum_{\alpha lm}\gamma _{\alpha lm}V_{\alpha lm}\rho
V_{\alpha lm}^{\dag }.  \label{JBipartito}
\end{equation}%
The measurement transformation [see Eq. (\ref{MOperator})] in the bipartite
Hilbert space becomes%
\begin{equation}
\mathbb{\hat{M}}\rho =\frac{\mathbb{\hat{J}}\rho }{\mathrm{Tr}_{sa}[\mathbb{%
\hat{J}}\rho ]}=\frac{\sum_{\alpha lm}\gamma _{\alpha lm}V_{\alpha lm}\rho
V_{\alpha lm}^{\dag }}{\left\{ \sum_{\alpha lm}\gamma _{\alpha lm}\mathrm{Tr}%
_{sa}[V_{\alpha lm}^{\dag }V_{\alpha lm}\rho ]\right\} }.  \label{MBipartito}
\end{equation}

The goal is to obtain a closed (stochastic) evolution for the system with
almost the same elements than in the Markovian case. The free parameters are
the rates $\gamma _{\alpha lm}.$ In order to have the same measurement
transformation than before [Eq. (\ref{MOperator})], for arbitrary bipartite
states $\rho _{sa}$ one must to demand the condition%
\begin{equation}
\mathrm{Tr}_{a}[\mathbb{\hat{M}}\rho _{sa}]=\mathcal{\hat{M}}[\rho _{s}],
\label{TracingM}
\end{equation}%
where evidently $\rho _{s}=\mathrm{Tr}_{a}[\rho _{sa}].$ There exist
different way of satisfying this condition. Here, for simplicity, we choose
the constraint%
\begin{equation}
\mathbb{\hat{M}}\rho _{sa}=\mathcal{\hat{M}}[\rho _{s}]\otimes \bar{\rho}%
_{a},  \label{Mapping}
\end{equation}%
where $\bar{\rho}_{a}$ is a particular ancilla density matrix. Notice that
after a measurement event, the system and ancilla become uncorrelated.
Trivially, this measurement transformation satisfy the previous condition
Eq. (\ref{TracingM}).

The conditional system dynamics between collision events can be written as
in Eq. (\ref{Conditional}), but now the unconditional propagator reads%
\begin{equation}
\mathcal{\hat{T}}(t-\tau )=\mathrm{Tr}_{a}[\exp [\mathbb{\hat{D}}(t-\tau )]%
\bar{\rho}_{a}].  \label{PropCondNoMarkov}
\end{equation}%
It arises from the partial trace over the ancilla system of the bipartite
conditional propagator $\mathbb{\hat{T}}(t-\tau )=\exp [\mathbb{\hat{D}}%
(t-\tau )],$ and the condition (\ref{Mapping}). The superoperator $\mathbb{%
\hat{D}}$ is%
\begin{equation}
\mathbb{\hat{D}}\rho =\mathbb{\hat{L}}_{0}\rho -\frac{1}{2}\sum_{\alpha
lm}\gamma _{\alpha lm}\{V_{\alpha lm}^{\dag }V_{\alpha lm},\rho \}_{+}.
\label{DBipartito}
\end{equation}

As we have chosen the stronger separability condition (\ref{Mapping}), the
propagator defined by $\mathcal{\hat{T}}(t)$ [Eq. (\ref{PropCondNoMarkov})]
is not only completely positive but also its time evolution is given by an
homogeneous equation. In fact, in a Laplace domain, $f(z)\equiv
\int_{0}^{\infty }dte^{-zt}f(t),$ Eq. (\ref{PropCondNoMarkov}) becomes $%
\mathcal{\hat{T}}(z)=\mathrm{Tr}_{a}[\frac{1}{z-\mathbb{\hat{D}}}\bar{\rho}%
_{a}].$ This expression can be rewritten as $\mathcal{\hat{T}}(z)=\{\mathrm{%
Tr}_{a}[(z-\mathbb{\hat{D}})^{-1}(z-\mathbb{\hat{D}})\bar{\rho}%
_{a}]\}^{-1}\times \{[\mathcal{\hat{T}}(z)]^{-1}\}^{-1}.$ Using in the curly
brackets that $M^{-1}\times N^{-1}=(N\times M)^{-1},$ where $M$ and $N$ are
arbitrary matrices, it follows $\mathcal{\hat{T}}(z)=\{[\mathcal{\hat{T}}%
(z)]^{-1}(z\mathrm{Tr}_{a}[(z-\mathbb{\hat{D}})^{-1}\bar{\rho}_{a}]-\mathrm{%
Tr}_{a}[(z-\mathbb{\hat{D}})^{-1}\mathbb{\hat{D}}\bar{\rho}_{a}])\}^{-1},$
which in turn leads to the expression $\mathcal{\hat{T}}(z)=[z-\mathcal{\hat{%
D}}(z)]^{-1},$ where the\ system superoperator $\mathcal{\hat{D}}(z)$ is%
\begin{equation}
\mathcal{\hat{D}}(z)=\Big{\{}\mathrm{Tr}_{a}\Big{[}\frac{1}{z-\mathbb{\hat{D}%
}}\bar{\rho}_{a}\Big{]}\Big{\}}^{-1}\mathrm{Tr}_{a}\Big{[}\frac{1}{z-\mathbb{%
\hat{D}}}\mathbb{\hat{D}}\bar{\rho}_{a}\Big{]}.  \label{DLaplace}
\end{equation}%
Hence, in the time domain we get%
\begin{equation}
\frac{d}{dt}\mathcal{\hat{T}}(t)=\int_{0}^{t}dt^{\prime }\mathcal{\hat{D}}%
(t-t^{\prime })\mathcal{\hat{T}}(t^{\prime }),  \label{d/dtT(t)}
\end{equation}%
where the memory superoperator $\mathcal{\hat{D}}(t)$ is defined by its
Laplace transform (\ref{DLaplace}). We notice that in the Markovian case $%
\mathcal{\hat{T}}(t)=\exp [t\mathcal{\hat{D}}]$ [see Eq. (\ref{Texponencial}%
)] implying the local in time evolution $(d/dt)\mathcal{\hat{T}}(t)=\mathcal{%
\hat{D}\hat{T}}(t).$ Thus, in the present approach the conditional evolution
between measurements events becomes non-local in time. This property also
implies that in general, even for pure initial conditions $\left\vert \Psi
\right\rangle ,$ the conditional evolution cannot be discomposed in pure
states \cite{plenio,carmichaelbook,breuerbook}, that is,%
\begin{equation}
\mathcal{\hat{T}}(t)(\left\vert \Psi \right\rangle \left\langle \Psi
\right\vert )\neq \left\vert \Psi (t)\right\rangle \left\langle \Psi
(t)\right\vert .  \label{NoSWF}
\end{equation}

Under the assumption Eq. (\ref{Mapping}), the previous analysis demonstrate
that it is possible to obtain a closed evolution for the system dynamics. It
remains to determine the statistics of the measurement events. As the
bipartite dynamics is Markovian, here we also have a well defined survival
probability [see Eq. (\ref{SurMarkov})]. By using Eq. (\ref{Mapping}), it is
possible to write 
\begin{subequations}
\label{SurvivalNoMarkov}
\begin{eqnarray}
P_{0}(t-\tau |\rho _{\tau }) &=&\mathrm{Tr}_{sa}[\exp [(t-\tau )\mathbb{\hat{%
D}}]\rho _{\tau }\otimes \bar{\rho}_{a}], \\
&=&\mathrm{Tr}_{s}[\mathcal{\hat{T}}(t-\tau )\rho _{\tau }],
\end{eqnarray}%
where $\mathcal{\hat{T}}(t)$ is given by Eq. (\ref{PropCondNoMarkov}).
Notice that $\rho _{\tau }$ is a system state. Furthermore, this expression
has the same structure than Eq. (\ref{SurMarkov}). The definition of the
conditional propagator $\mathcal{\hat{T}}(t)$ is the unique difference. The
corresponding waiting time distribution [Eq. (\ref{WaitMarkov})] here reads 
\end{subequations}
\begin{equation}
w(t-\tau |\rho _{\tau })=-\mathrm{Tr}_{sa}[\mathbb{\hat{D}}\exp [(t-\tau )%
\mathbb{\hat{D}}]\rho _{\tau }\otimes \bar{\rho}_{a}].
\end{equation}%
From Eq. (\ref{d/dtT(t)}) it follows the equivalent expression%
\begin{equation}
w(t-\tau |\rho _{\tau })=-\int_{0}^{t-\tau }dt^{\prime }\mathrm{Tr}_{s}[%
\mathcal{\hat{D}}(t-\tau -t^{\prime })\mathcal{\hat{T}}(t^{\prime })\rho
_{\tau }],  \label{waitingNoMarkov}
\end{equation}%
which leads to a natural non-Markovian generalization of Eq. (\ref%
{WaitMarkov}). On the other hand, the conditional waiting time distribution,
Eq. (\ref{WaitConditional}), here becomes%
\begin{equation}
w_{c}(t-\tau |\rho _{\tau })=\frac{-\int_{0}^{t-\tau }dt^{\prime }\mathrm{Tr}%
_{s}[\mathcal{\hat{D}}(t-\tau -t^{\prime })\mathcal{\hat{T}}(t^{\prime
})\rho _{\tau }]}{\mathrm{Tr}_{s}[\mathcal{\hat{T}}(t-\tau )\rho _{\tau }]}.
\label{WaitCondNoMarkov}
\end{equation}

\subsection{Stochastic dynamics}

As in the Markovian case, the previous objects [Eqs. (\ref{Mapping}), (\ref%
{d/dtT(t)}), and (\ref{SurvivalNoMarkov})] completely define the system
realizations associated to the measurement process. Therefore, the algorithm
associated to Eq. (\ref{PEstocastica}) remains exactly the same. The unique
modification is the definition of the propagator $\mathcal{\hat{T}}(t),$
which in turn modify the conditional dynamics as well as the measurement
events statistics.

On the other hand, the infinitesimal time step algorithm defined by Eq. (\ref%
{DeltaPDefinicion}) can also be applied. Nevertheless, in contrast to Eq.(%
\ref{DeltaPMarkov}), here it is not possible to write a simple expression
for $\Delta P$ neither in terms of $\rho _{\mathrm{st}}^{s}(t)$ or its
history [see Eq. (\ref{WaitCondNoMarkov})]. Therefore, in this generalized
non-Markovian approach the infinitesimal algorithm, while can be formally
implemented, it does not provide an efficient numerical simulation method
neither it has a simple physical interpretation.

\subsection{Symmetries of the bipartite dynamics}

It remains to demonstrate that in fact there exist different bipartite
Lindblad equations that allow to fulfill the condition (\ref{Mapping}),
where the bipartite measurement transformation is given by Eq. (\ref%
{MBipartito}). From Eq. (\ref{VAlfaLM}), it can be written as%
\begin{equation}
\mathbb{\hat{M}}\rho =\frac{\sum_{\alpha lm}\gamma _{\alpha lm}V_{\alpha
}\left\langle a_{m}\right\vert \rho \left\vert a_{m}\right\rangle V_{\alpha
}^{\dag }\otimes \left\vert a_{l}\right\rangle \left\langle a_{l}\right\vert 
}{\left\{ \sum_{\alpha lm}\gamma _{\alpha lm}\mathrm{Tr}_{s}[V_{\alpha
}\left\langle a_{m}\right\vert \rho \left\vert a_{m}\right\rangle V_{\alpha
}^{\dag }]\right\} }.  \label{M_General}
\end{equation}%
The result of calculating $\mathrm{Tr}_{a}[\mathbb{\hat{M}}\rho ]$ can only
be written in terms of $\mathcal{\hat{M}}$ [Eq. (\ref{MOperator})] if $%
\gamma _{\alpha lm}=\gamma _{\alpha l}d_{m},$ where $d_{m}$\ is an arbitrary
dimensionless coefficient. Eq. (\ref{M_General}) becomes%
\begin{equation}
\mathbb{\hat{M}}\rho =\frac{\sum_{\alpha m}\gamma _{\alpha }V_{\alpha
}d_{m}\left\langle a_{m}\right\vert \rho \left\vert a_{m}\right\rangle
V_{\alpha }^{\dag }\otimes \bar{\rho}_{a}^{\alpha }}{\left\{ \sum_{\alpha
m}\gamma _{\alpha }\mathrm{Tr}_{s}[V_{\alpha }d_{m}\left\langle
a_{m}\right\vert \rho \left\vert a_{m}\right\rangle V_{\alpha }^{\dag
}]\right\} },  \label{M_Discordia}
\end{equation}%
where $\bar{\rho}_{a}^{\alpha }\equiv \sum_{l}(\gamma _{\alpha l}/\gamma
_{\alpha })\left\vert a_{l}\right\rangle \left\langle a_{l}\right\vert ,$
and $\gamma _{\alpha }\equiv \sum_{l}\gamma _{\alpha l}.$ With the operators
definitions (\ref{ValfaRenewal}) and (\ref{ValfaNoRenewal}), Eq. (\ref%
{M_Discordia}) can satisfy the weaker condition (\ref{TracingM}).
Nevertheless, the resulting bipartite state is a classical correlated one
(with vanishing discord). For satisfying the separability condition (\ref%
{Mapping}), which leads to the homogeneous dynamics (\ref{d/dtT(t)}), the
states $\bar{\rho}_{a}^{\alpha }$ must not to depend on index $\alpha .$
Hence, we demand $\gamma _{\alpha l}=\gamma _{\alpha }c_{l},$ where $c_{l}$
is also an arbitrary dimensionless coefficient. The rates $\gamma _{\alpha
lm}$ become%
\begin{equation}
\gamma _{\alpha lm}=\gamma _{\alpha }c_{l}d_{m},\ \ \ \ \ \ \
\sum_{l}c_{l}=1,  \label{rateCondition}
\end{equation}%
which from Eq. (\ref{M_General}) leads to%
\begin{equation}
\mathbb{\hat{M}}\rho =\mathcal{\hat{M}}\big{[}\sum_{m}d_{m}\left\langle
a_{m}\right\vert \rho \left\vert a_{m}\right\rangle \big{]}\otimes \bar{\rho}%
_{a}.  \label{McasiListo}
\end{equation}%
The ancilla resetting state $\bar{\rho}_{a}$\ is%
\begin{equation}
\bar{\rho}_{a}=\sum_{l}c_{l}\left\vert a_{l}\right\rangle \left\langle
a_{l}\right\vert .  \label{AncillaReseting}
\end{equation}%
For simplicity, we assumed $\sum_{l}c_{l}=1.$ If this condition is not meet,
it can always be satisfied by a renormalization of the measurement rates, $%
\gamma _{\alpha }\rightarrow \gamma _{\alpha }/\sum_{l}c_{l}.$

The expression (\ref{rateCondition}) can be read as a symmetry condition on
the bipartite Lindblad evolution Eq. (\ref{LindbladBipartito}). It leads to
Eq. (\ref{McasiListo}), which does not recover explicitly Eq. (\ref{Mapping}%
). The fulfilment of this constraint can be achieved by choosing different
set of values for the coefficients $d_{m},$ which depend on the specific
structure of $\mathcal{\hat{M}}.$

\subsubsection{Renewal case}

When the measurement transformation $\mathcal{\hat{M}}$ leads to a renewal
process, Eqs. (\ref{ValfaRenewal}) and (\ref{Mrenewal}), independently of
the coefficients $d_{m}$ it follows $\mathcal{\hat{M}}[\sum_{m}d_{m}\left%
\langle a_{m}\right\vert \rho \left\vert a_{m}\right\rangle ]=\bar{\rho}%
_{s}. $ Therefore, Eq. (\ref{McasiListo}) leads to%
\begin{equation}
\mathbb{\hat{M}}\rho =\bar{\rho}_{s}\otimes \bar{\rho}_{a}.
\label{ResettingBipartito}
\end{equation}%
Evidently this expression satisfies the condition (\ref{Mapping}).
Furthermore, it says us that the stochastic dynamics developing in the
bipartite $S-A$ Hilbert space is also a renewal measurement process.

Similarly to the Markovian case, after the first detection event the
statistic of the time interval between consecutive events is defined by a
unique survival probability%
\begin{equation}
P_{0}(t)=\mathrm{Tr}_{s}[\mathcal{\hat{T}}(t)\bar{\rho}_{s}],
\label{SurNoMarkRenewal}
\end{equation}%
or equivalently a unique waiting time distribution%
\begin{equation}
w(t)=-\int_{0}^{t}dt^{\prime }\mathrm{Tr}_{s}[\mathcal{\hat{D}}(t-t^{\prime
})\mathcal{\hat{T}}(t^{\prime })\bar{\rho}_{s}].  \label{WaiterNoMarkRenewal}
\end{equation}%
These expressions follows from Eqs. (\ref{SurvivalNoMarkov}) and (\ref%
{waitingNoMarkov}) after introducing the resetting property defined by Eq. (%
\ref{ResettingBipartito}). They generalize the Markovian expressions (\ref%
{MarkovRenewal}).

\subsubsection{Non-renewal case}

When the measurement transformation $\mathcal{\hat{M}}$\ does not lead to a
renewal process [Eqs. (\ref{ValfaNoRenewal}) and (\ref{MNoRenewal})], the
coefficients $d_{m}$\ can not be arbitrary. In fact, the only way of
satisfying the condition (\ref{Mapping}) is by choosing $d_{m}=1$ (after a
rates renormalization we can also take $d_{m}$ equal to an arbitrary real
constant). As the states $\{\left\vert a_{m}\right\rangle \}$ are a complete
basis of the ancilla Hilbert space, for any bipartite state $\rho _{sa}$ it
follows $\sum_{m}\left\langle a_{m}\right\vert \rho _{sa}\left\vert
a_{m}\right\rangle =\mathrm{Tr}_{a}[\rho _{sa}]=\rho _{s}.$ Thus, Eq. (\ref%
{McasiListo}) recovers Eq. (\ref{Mapping}),%
\begin{equation}
\mathbb{\hat{M}}\rho =\mathcal{\hat{M}}[\rho _{s}]\otimes \bar{\rho}_{a}.
\end{equation}%
Notice that this result is valid for both the non-renewal and renewal cases.
Nevertheless, the condition $d_{m}=1$ is only \textquotedblleft
necessary\textquotedblright\ in the former case. The symmetry condition on
the rates $\gamma _{\alpha lm}$ [Eq. (\ref{rateCondition})] then reads%
\begin{equation}
\gamma _{\alpha lm}=\gamma _{\alpha }c_{l},\ \ \ \ \ \ \ \sum_{l}c_{l}=1.
\label{rateConditionNoRenewal}
\end{equation}%
In contrast to the renewal case, here the measurement statistics remains
defined by the general expression Eqs. (\ref{SurvivalNoMarkov}) and (\ref%
{waitingNoMarkov}).

\subsection{Density matrix evolution}

Under the symmetry conditions defined by Eqs. (\ref{rateCondition}) and (\ref%
{rateConditionNoRenewal}) the stochastic dynamics of $\rho _{\mathrm{st}%
}^{s}(t)$ [Eq. (\ref{RhoSystemEstocastica})] has the same structure than in
the Markovian case. For both renewal and non-renewal measurement processes,
the main difference with the Markovian case is the conditional dynamics. It
remains to calculate the time evolution of the system density matrix $\rho
_{t}^{s}$, Eq. (\ref{RhoPartialSistema}). In Appendix B we perform this
calculus by averaging the realizations of $\rho _{\mathrm{st}}^{s}(t),$ that
is, from $\rho _{t}^{s}=\overline{\rho _{\mathrm{st}}^{s}(t)}.$ Here, using
an alternative procedure, the evolution of the system state is obtained from
the bipartite dynamics (\ref{LindbladBipartito}) by using that $\rho
_{t}^{s}=\mathrm{Tr}_{a}[\rho _{t}^{s}].$

For simplicity, we take a separable bipartite initial condition%
\begin{equation}
\rho _{0}^{sa}=\rho _{0}^{s}\otimes \bar{\rho}_{a},  \label{Inicial}
\end{equation}%
where $\rho _{0}^{s}$ is an arbitrary system state and $\bar{\rho}_{a}$ is
the ancilla resetting state defined by Eq. (\ref{AncillaReseting}). The
bipartite Lindblad evolution (\ref{LindbladBipartito}) can formally be
integrated as%
\begin{equation}
\rho _{t}^{sa}=\exp [\mathbb{\hat{D}}t]\rho _{0}^{sa}+\int_{0}^{t}dt^{\prime
}\exp [\mathbb{\hat{D}}(t-t^{\prime })]\mathbb{\hat{J}}[\rho _{t^{\prime
}}^{sa}].  \label{RhoSAIntegrada}
\end{equation}%
The superoperators $\mathbb{\hat{J}}$ and $\mathbb{\hat{D}}$\ were defined
in Eqs. (\ref{JBipartito}) and (\ref{DBipartito}) respectively. By using the
rates condition Eq. (\ref{rateCondition}) and the operator definition (\ref%
{VAlfaLM}), we get%
\begin{equation}
\mathbb{\hat{J}}[\rho _{t}^{sa}]=\sum_{\alpha }\gamma _{\alpha }V_{\alpha }%
\mathrm{O}[\rho _{t}^{sa}]V_{\alpha }^{\dag }\otimes \bar{\rho}_{a}.
\end{equation}%
For shortening the notation we defined the superoperator%
\begin{equation}
\mathrm{O}[\rho _{t}^{sa}]\equiv \sum_{m}d_{m}\left\langle a_{m}\right\vert
\rho _{t}^{sa}\left\vert a_{m}\right\rangle .  \label{Ouhh}
\end{equation}%
Taking the partial trace over the ancilla degrees of freedom, Eq. (\ref%
{RhoSAIntegrada}) leads to%
\begin{equation}
\rho _{t}^{s}=\mathcal{\hat{T}}(t)\rho _{0}^{s}+\int_{0}^{t}dt^{\prime }%
\mathcal{\hat{T}}(t-t^{\prime })\sum_{\alpha }\gamma _{\alpha }V_{\alpha }%
\mathrm{O}[\rho _{t^{\prime }}^{sa}]V_{\alpha }^{\dag },
\end{equation}%
which in turn, from Eq. (\ref{d/dtT(t)}), allows us to write%
\begin{equation}
\frac{d\rho _{t}^{s}}{dt}=\int_{0}^{t}dt^{\prime }\mathcal{\hat{D}}%
(t-t^{\prime })\rho _{t^{\prime }}^{s}+\sum_{\alpha }\gamma _{\alpha
}V_{\alpha }\mathrm{O}[\rho _{t}^{sa}]V_{\alpha }^{\dag }.
\label{MasterPrevia}
\end{equation}

If all $d_{m}=1,$ it follows $\mathrm{O}[\rho _{t}^{sa}]=\rho _{t}^{s}.$
Hence, from Eq. (\ref{MasterPrevia}) we get the closed density matrix
evolution%
\begin{equation}
\frac{d\rho _{t}^{s}}{dt}=\int_{0}^{t}dt^{\prime }\mathcal{\hat{D}}%
(t-t^{\prime })\rho _{t^{\prime }}^{s}+\sum_{\alpha }\gamma _{\alpha
}V_{\alpha }\rho _{t}^{s}V_{\alpha }^{\dag }.  \label{LocalNonLocal}
\end{equation}%
Notice that this evolution contains both convoluted as well as local in time
contributions. It is valid for both, renewal and non-renewal measurement
processes. On the other hand, in the case of renewal processes the
coefficients $d_{m}$ may be arbitrary and the previous expression does not
apply. By using the specific form of the operators $V_{\alpha }$ [Eq. (\ref%
{ValfaRenewal})] it follows $\sum_{\alpha }\gamma _{\alpha }V_{\alpha }%
\mathrm{O}[\rho _{t}^{sa}]V_{\alpha }^{\dag }=\bar{\rho}_{s}\gamma
\left\langle u\right\vert \mathrm{O}[\rho _{t}^{sa}]\left\vert
u\right\rangle ,$ where $\gamma =\sum_{\alpha }\gamma _{\alpha }$ and the
system resetting state $\bar{\rho}_{s}$\ is defined by Eq. (\ref{Mrenewal}).
By using in Eq. (\ref{MasterPrevia}) that $(d/dt)\mathrm{Tr}_{s}[\rho
_{t}^{s}]=0,$ it follows $\gamma \left\langle u\right\vert \mathrm{O}[\rho
_{t}^{sa}]\left\vert u\right\rangle =-\int_{0}^{t}dt^{\prime }\mathrm{Tr}%
_{s}[\mathcal{\hat{D}}(t-t^{\prime })\rho _{t^{\prime }}^{s}],$ implying the
closed density matrix evolution%
\begin{equation}
\frac{d\rho _{t}^{s}}{dt}=\int_{0}^{t}dt^{\prime }\mathcal{\hat{D}}%
(t-t^{\prime })\rho _{t^{\prime }}^{s}-\bar{\rho}_{s}\int_{0}^{t}dt^{\prime }%
\mathrm{Tr}_{s}[\mathcal{\hat{D}}(t-t^{\prime })\rho _{t^{\prime }}^{s}].
\label{MasterRenewal}
\end{equation}%
In the present approach, this expression\ correspond to the more general
master equation consistent with a renewal measurement process. Notice that
Eq. (\ref{LocalNonLocal}) is a particular case of this more general
expression. By comparing both equations, we realize that it applies when $%
\gamma \left\langle u\right\vert \rho _{t}^{s}\left\vert u\right\rangle
=\int_{0}^{t}dt^{\prime }\mathrm{Tr}_{s}[\mathcal{\hat{D}}(t-t^{\prime
})\rho _{t^{\prime }}^{s}].$

\subsection{Arbitrary master equations}

Eqs. (\ref{LocalNonLocal}) and (\ref{MasterRenewal}) are one of the central
results of this section. They correspond to master equations that admit an
unravelling in terms of an ensemble of trajectories associated to a
continuous measurement action defined by the set of operators $\{V_{\alpha
}\}.$ Eq. (\ref{LocalNonLocal}) is valid for both renewal and non-renewal
measurement processes [see Eqs. (\ref{ValfaRenewal}) and (\ref%
{ValfaNoRenewal}) respectively] while Eq. (\ref{MasterRenewal}) is only
valid for renewal processes [Eq. (\ref{ValfaRenewal})]. Now we ask about
which conditions an arbitrary non-Markovian master equation must to satisfy
to admit the non-Markovian unravelling defined previously.

One condition is the possibility of rewriting the master equation with the
structure defined by Eqs. (\ref{LocalNonLocal}) or (\ref{MasterRenewal}). On
the other hand, the ensemble representation can only be assigned if the
memory superoperator $\mathcal{\hat{D}}(t)$ through the relation $(d/dt)%
\mathcal{\hat{T}}(t)=\int_{0}^{t}dt^{\prime }\mathcal{\hat{D}}(t-t^{\prime })%
\mathcal{\hat{T}}(t^{\prime })$ [Eq. (\ref{d/dtT(t)})] defines a well
behaved survival probability $P_{0}(t-\tau |\rho )=\mathrm{Tr}_{s}[\mathcal{%
\hat{T}}(t-\tau )\rho ]$ [Eq. (\ref{SurvivalNoMarkov})] for
\textquotedblleft arbitrary\textquotedblright\ system states $\rho .$ A well
behaved survival probability means that it is a decaying function, that is,
for arbitrary times $\tau <t_{1}<t_{2},$ it must to satisfy $%
P_{0}(t_{2}-\tau |\rho )\leq P_{0}(t_{1}-\tau |\rho ),$ implying%
\begin{equation}
\mathrm{Tr}_{s}[\mathcal{\hat{T}}(t_{2})\rho ]\leq \mathrm{Tr}_{s}[\mathcal{%
\hat{T}}(t_{1})\rho ],\ \ \ \ \ \ \ \ t_{1}<t_{2}.
\label{PropagatorConditionSur}
\end{equation}%
Taking into account that the realizations can be determine from $%
P_{0}(t|\rho ),$ the fulfillment of the previous inequality guarantees the
possibility of assigning a non-Markovian unravelling to a master equation
with the structure (\ref{LocalNonLocal}) or (\ref{MasterRenewal}).

\section{Non-Markovian renewal two-level transitions}

Here, we work out an example that explicitly shows the consistence of the
previous results. Both the system of interest and the ancilla are two-level
systems. Their states are denoted $\left\vert \pm \right\rangle ,$ and $%
\{\left\vert 1\right\rangle ,\left\vert 2\right\rangle \}$ respectively. The
Markovian dynamic of the bipartite state $\rho _{t}^{sa}$\ [Eq. (\ref%
{LindbladBipartito})] here reads%
\begin{equation}
\frac{d}{dt}\rho _{t}^{sa}=-\frac{i}{\hbar }[H_{0},\rho _{t}^{sa}]+(\gamma 
\mathcal{C}[\sigma _{11}]+\gamma ^{\prime }\mathcal{C}[\sigma _{21}])\rho
_{t}^{sa}.  \label{LindbladTLS}
\end{equation}%
The bipartite Hamiltonian contribution is defined by the operator%
\begin{equation}
H_{0}=\hbar \Omega \sigma _{x}\otimes \sigma _{x},
\end{equation}%
where $\sigma _{x}$ is the $x$-Pauli matrix in the basis of each Hilbert
space. The remaining Lindblad contributions [Eq. (\ref{LindbladDEF})] with
rates $\gamma $\ and $\gamma ^{\prime }$ are defined by the operators%
\begin{equation}
\sigma _{11}=\sigma \otimes \left\vert 1\right\rangle \left\langle
1\right\vert ,\ \ \ \ \ \ \ \ \ \ \sigma _{12}=\sigma \otimes \left\vert
1\right\rangle \left\langle 2\right\vert .
\end{equation}%
The lowering system operator is defined as $\sigma =\left\vert
-\right\rangle \left\langle +\right\vert .$

Notice that $\sigma _{11}$ leads to system transitions between the upper and
lower states $\left\vert +\right\rangle \rightsquigarrow \left\vert
-\right\rangle $\ that can only happen when the ancilla is in the state $%
\left\vert 1\right\rangle .$ In addition, $\sigma _{12}$ leads to the same
system transitions but in this case they simultaneously occur with the
ancilla transition $\left\vert 2\right\rangle \rightsquigarrow \left\vert
1\right\rangle .$ Thus, the dissipative dynamic drives the system to its
ground states. On the other hand, the unitary evolution can excite the
system to its upper state. In consequence, the interplay between both
contributions leads to successive system transitions $\left\vert
+\right\rangle \rightsquigarrow \left\vert -\right\rangle .$ Each transition
can be associated with a recording event in the measurement apparatus.

It is simple to check that Eq. (\ref{LindbladTLS}) has the structure defined
by Eqs. (\ref{LindbladBipartito})-(\ref{VAlfaLM}), and also fulfill the
symmetry condition Eq. (\ref{rateCondition}). Consistently with the previous
analysis, the superoperator $\mathbb{\hat{J}}$ [Eq. (\ref{JBipartito})] is
defined as%
\begin{equation}
\mathbb{\hat{J}}\rho =\gamma \sigma _{11}\rho \sigma _{11}^{\dag }+\gamma
^{\prime }\sigma _{12}\rho \sigma _{12}^{\dag },
\end{equation}%
leading to the expression%
\begin{equation*}
\mathbb{\hat{J}}\rho =(\gamma \left\langle +1\right\vert \rho \left\vert
+1\right\rangle +\gamma ^{\prime }\left\langle +2\right\vert \rho \left\vert
+2\right\rangle )\left\vert -\right\rangle \left\langle -\right\vert \otimes
\left\vert 1\right\rangle \left\langle 1\right\vert .
\end{equation*}%
From here, the measurement transformation [Eq. (\ref{MBipartito})]
associated to each event reads%
\begin{equation}
\mathbb{\hat{M}}\rho =\left\vert -\right\rangle \left\langle -\right\vert
\otimes \left\vert 1\right\rangle \left\langle 1\right\vert .
\label{ResetingExample}
\end{equation}%
Therefore, the state after a detection is independent of the previous
bipartite state $\rho ,$ which in turn implies that the measurement process
is a renewal one [see Eq. (\ref{ResettingBipartito})]. The bipartite
conditional dynamics between events is defined by the superoperator [Eq. (%
\ref{DBipartito})]%
\begin{equation}
\mathbb{\hat{D}}\rho =-\frac{i}{\hbar }[H_{0},\rho ]-\frac{1}{2}\{(\gamma
\sigma _{11}^{\dag }\sigma _{11}+\gamma ^{\prime }\sigma _{12}^{\dag }\sigma
_{12}),\rho \}_{+}.
\end{equation}%
In order to obtain simple analytical expressions from now on we analyze the
case $\gamma ^{\prime }=\gamma .$ Notice that it is also possible to take $%
\gamma ^{\prime }=0$ with $\gamma >0,$ or $\gamma =0$ with $\gamma ^{\prime
}>0.$

The conditional propagator $\mathcal{\hat{T}}(t)$\ [Eq. (\ref%
{PropCondNoMarkov})] can be defined when acting on an arbitrary initial
condition $\rho .$ By defining the state $\tilde{\rho}_{t}=\mathcal{\hat{T}}%
(t)\rho ,$ the time evolution of $\mathcal{\hat{T}}(t)$ [Eq. (\ref{d/dtT(t)}%
)] can be written in terms of the matrix elements%
\begin{equation}
\tilde{p}_{t}^{\pm }\equiv \left\langle \pm \right\vert \tilde{\rho}%
_{t}\left\vert \pm \right\rangle ,\ \ \ \ \ \tilde{c}_{t}^{\pm }\equiv
\left\langle \pm \right\vert \tilde{\rho}_{t}\left\vert \mp \right\rangle .
\label{MatrixElementsConditional}
\end{equation}%
For the populations we get 
\begin{subequations}
\label{PoblacionCondicional}
\begin{eqnarray}
\frac{d\tilde{p}_{t}^{+}}{dt} &=&-\int_{0}^{t}dt^{\prime }k_{t-t^{\prime
}}^{+}\tilde{p}_{t^{\prime }}^{+}+\int_{0}^{t}dt^{\prime }k_{t-t^{\prime
}}^{-}\tilde{p}_{t^{\prime }}^{-}, \\
\frac{d\tilde{p}_{t}^{-}}{dt} &=&-\int_{0}^{t}dt^{\prime }k_{t-t^{\prime
}}^{-}\tilde{p}_{t^{\prime }}^{-}+(1-\tilde{\delta})\int_{0}^{t}dt^{\prime
}k_{t-t^{\prime }}^{+}\tilde{p}_{t^{\prime }}^{+}.\ \ \ \ \ \ \ 
\end{eqnarray}%
Here, the constant $\tilde{\delta}$ must be taken as $\tilde{\delta}%
\rightarrow 1.$ Thus, the last term does not contribute. The memory kernels
are 
\end{subequations}
\begin{equation}
k_{t}^{+}=\gamma \delta (t)+\frac{\Omega ^{2}}{2}e^{-t\gamma /2},\ \ \ \ \ \
\ \ k_{t}^{-}=\frac{\Omega ^{2}}{2}e^{-t\gamma /2}.
\end{equation}%
The coherence evolve as%
\begin{equation}
\frac{d\tilde{c}_{t}^{\pm }}{dt}=-\int_{0}^{t}dt^{\prime }\tilde{k}%
_{t-t^{\prime }}\tilde{c}_{t^{\prime }}^{\pm }+\int_{0}^{t}dt^{\prime }%
\breve{k}_{t-t^{\prime }}\tilde{c}_{t^{\prime }}^{\mp },
\label{CoherenciaCondicional}
\end{equation}%
where the kernels $\tilde{k}_{t}$\ and$\ \breve{k}_{t}$ are%
\begin{equation}
\tilde{k}_{t}=\frac{\gamma }{2}\delta (t)+\frac{\Omega ^{2}}{4}%
(1+e^{-t\gamma }),\ \ \ \ \ \ \ \ \breve{k}_{t}=\frac{\Omega ^{2}}{4}%
(1+e^{-t\gamma }).
\end{equation}%
Due to the symmetries of the problem, populations and coherences evolve
independently each of the others.

Eqs. (\ref{PoblacionCondicional}) and (\ref{CoherenciaCondicional}) can be
solved in a Laplace domain. The survival probability [Eq. (\ref%
{SurvivalNoMarkov})] reads $P_{0}(t|\rho )=\mathrm{Tr}_{s}[\mathcal{\hat{T}}%
(t)\rho ]=\mathrm{Tr}_{s}[\tilde{\rho}_{t}]=\tilde{p}_{t}^{+}+\tilde{p}%
_{t}^{-}.$ We get%
\begin{eqnarray}
P_{0}(t|\rho ) &=&\mathrm{Tr}_{s}[\rho ]e^{-\frac{\gamma t}{2}}\Big{[}\Big{(}%
\frac{\gamma }{2\nu }\Big{)}^{2}\cosh (\nu t)-\Big{(}\frac{\Omega }{\nu }%
\Big{)}^{2}\Big{]}  \notag \\
&&-\mathrm{Tr}_{s}[\sigma _{z}\rho ]e^{-\frac{\gamma t}{2}}\Big{[}\frac{%
\gamma }{2\nu }\sinh (\nu t)\Big{]},  \label{SurvivalAnalitica}
\end{eqnarray}%
where the \textquotedblleft frequency\textquotedblright\ $\nu $ reads%
\begin{equation}
\nu =\sqrt{(\gamma /2)^{2}-\Omega ^{2}}.
\end{equation}%
In Eq. (\ref{SurvivalAnalitica}) the dependence on the system state $\rho $
is given $\mathrm{Tr}_{s}[\rho ]$ and $\mathrm{Tr}_{s}[\sigma _{z}\rho ],$
where $\sigma _{z}$ is the $z$-Pauli Matrix. Using the normalization of $%
\rho $ it follows $\mathrm{Tr}_{s}[\rho ]=1,$ while $\mathrm{Tr}_{s}[\sigma
_{z}\rho ]=\left\langle +\right\vert \rho \left\vert +\right\rangle
-\left\langle -\right\vert \rho \left\vert -\right\rangle .$ Therefore, $%
P_{0}(t|\rho )$ only depends on the populations of $\rho .$

In Fig. 1 we plotted $P_{0}(t|\rho )$ and its associated waiting time
distribution, $w(t|\rho )=-(d/dt)P_{0}(t|\rho )$ [Eq. (\ref{waitingNoMarkov}%
)] for different initial states $\rho .$ In Figs. 1(a) and (b) we took $\rho
=\left\vert y_{-}\right\rangle \left\langle y_{-}\right\vert ,$ where $%
\left\vert y_{-}\right\rangle $ is an eigenvector of $\sigma _{y}$ with
eigenvalue minus one, $\left\vert y_{-}\right\rangle =(1/\sqrt{2}%
)(\left\vert +\right\rangle -i\left\vert -\right\rangle ).$ In Figs. 1(c)
and (d) the initial state is $\rho =\left\vert -\right\rangle \left\langle
-\right\vert ,$ that is, the resetting state after a detection event [see
Eq. (\ref{ResetingExample})]. Hence, these objects, after the first
measurement event, completely define the measurement statistics [Eqs. (\ref%
{SurNoMarkRenewal}) and (\ref{WaiterNoMarkRenewal})]. Consistently with Eq. (%
\ref{PropagatorConditionSur}), for both initial conditions the survival
probabilities as a function of time are decaying functions. On the other
hand, while $\lim_{t\rightarrow 0}w(t|\left\vert y_{-}\right\rangle
\left\langle y_{-}\right\vert )\neq 0,$ given that $\lim_{t\rightarrow
0}w(t)=0,$ an antibunching phenomenon \cite{breuerbook,carmichaelbook}
characterize the renewal measurement process.

The survival probability allows to generate the random time intervals
between detection events. On the other hand, the matrix elements of $\tilde{%
\rho}_{t}=\mathcal{\hat{T}}(t)\rho $ [Eq. (\ref{MatrixElementsConditional})]
also allows to obtain the corresponding normalized conditional evolution $%
\mathcal{\hat{T}}(t)\rho /\mathrm{Tr}_{s}[\mathcal{\hat{T}}(t)\rho ]=\tilde{%
\rho}_{t}/\mathrm{Tr}_{s}[\tilde{\rho}_{t}].$ In each jump, the measurement
transformation $\rho \rightarrow \mathcal{M}\rho =\left\vert -\right\rangle
\left\langle -\right\vert $ applies [see Eq. (\ref{ResetingExample})]. These
elements completely define the ensemble of trajectories associated to the
stochastic density matrix $\rho _{\mathrm{st}}^{s}(t)$ (see Sec. III-B). 
\begin{figure}[tbp]
\includegraphics[bb=50 572 693 1100,width=8.5cm]{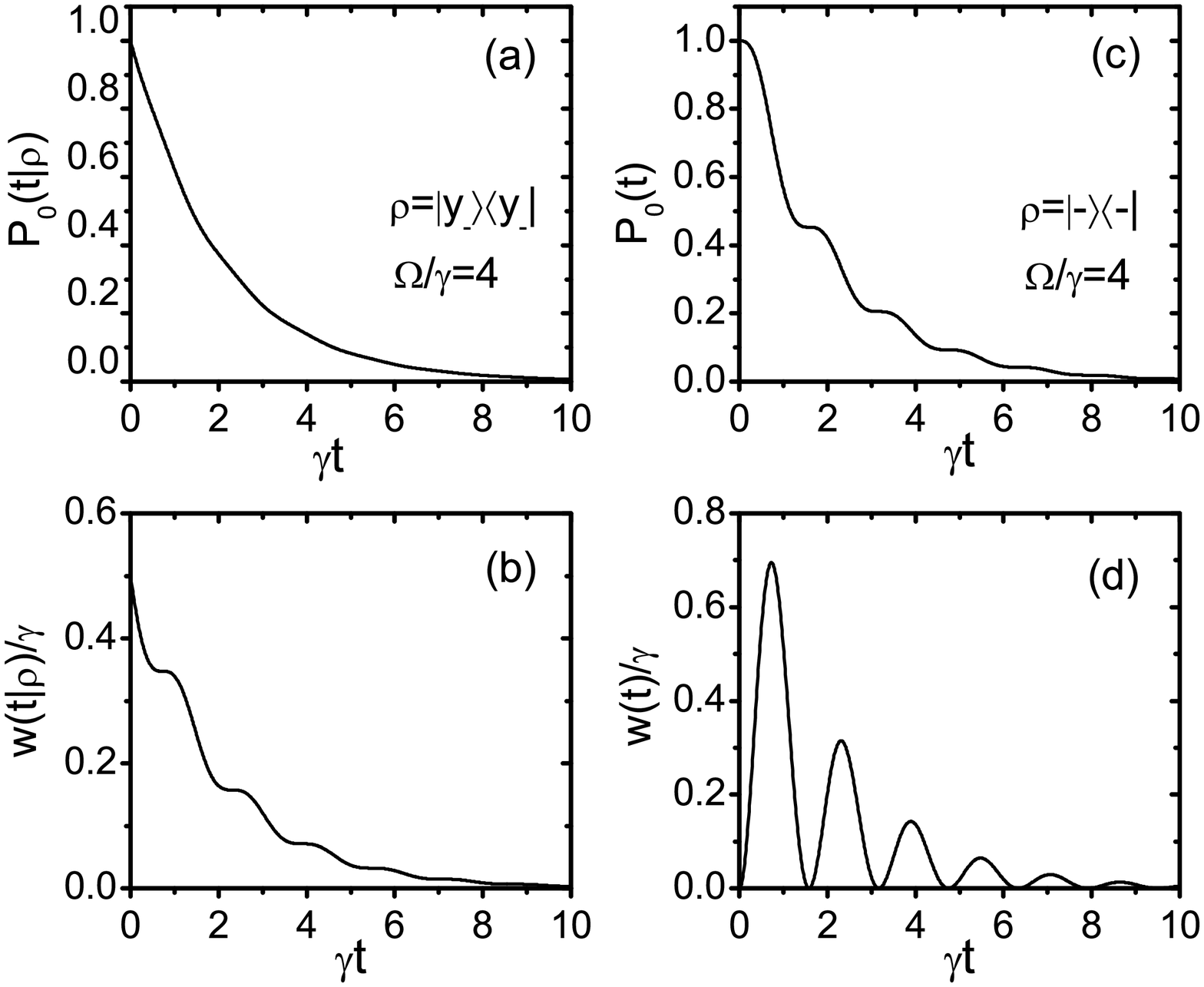}
\caption{Survival probability $P_{0}(t|\protect\rho )$ [Eq. (\protect\ref%
{SurvivalAnalitica})] and its associated waiting time distribution $w(t|%
\protect\rho )=-(d/dt)P_{0}(t|\protect\rho )$ for different initial
conditions. In (a) and (b), $\protect\rho =\left\vert y_{-}\right\rangle
\left\langle y_{-}\right\vert ,$ where $\left\vert y_{-}\right\rangle =(1/%
\protect\sqrt{2})(\left\vert +\right\rangle -i\left\vert -\right\rangle ).$
In (c) and (d), $\protect\rho =\left\vert -\right\rangle \left\langle
-\right\vert ,$ which correspond to the resetting state defined by Eq. (%
\protect\ref{ResetingExample}). In all cases, the parameters satisfies $%
\Omega /\protect\gamma =4.$}
\end{figure}

In Fig. 2, a realization of $\rho _{\mathrm{st}}^{s}(t)$ is showed through
the matrix elements $\left\langle +\right\vert \rho _{\mathrm{st}%
}^{s}(t)\left\vert +\right\rangle $ [upper population, Fig. 2(a)] and $%
\left\langle +\right\vert \rho _{\mathrm{st}}^{s}(t)\left\vert
-\right\rangle $ [coherence, Fig. 2(b)]. The initial state is $\rho _{%
\mathrm{st}}^{s}(0)=\left\vert y_{-}\right\rangle \left\langle
y_{-}\right\vert .$ In the behavior of $\left\langle +\right\vert \rho _{%
\mathrm{st}}^{s}(t)\left\vert +\right\rangle $ it is possible to observe the
successive jumps, where the system state collapse to the resetting state $%
\left\vert -\right\rangle \left\langle -\right\vert ,$ or equivalently, $%
\left\langle +\right\vert \rho _{\mathrm{st}}^{s}(t)\left\vert
+\right\rangle \rightarrow 0.$The conditional interevent behavior is
periodic.

On the other hand, for the chosen initial condition the coherence $%
\left\langle +\right\vert \rho _{\mathrm{st}}^{s}(t)\left\vert
-\right\rangle $ does not have a real component. Hence, from Fig. 2(b) we
conclude that after the first event is dyes out. This property follows form
the resetting state defined by Eq. (\ref{ResetingExample}) and the fact that
the conditional evolution [Eqs. (\ref{PoblacionCondicional}) and (\ref%
{CoherenciaCondicional})] does not couple the populations and coherences of
the system. Notice that for the chosen parameters values an oscillatory
behavior characterize the conditional coherence dynamics. 
\begin{figure}[tbp]
\includegraphics[bb=42 585 705 1130,width=8.5cm]{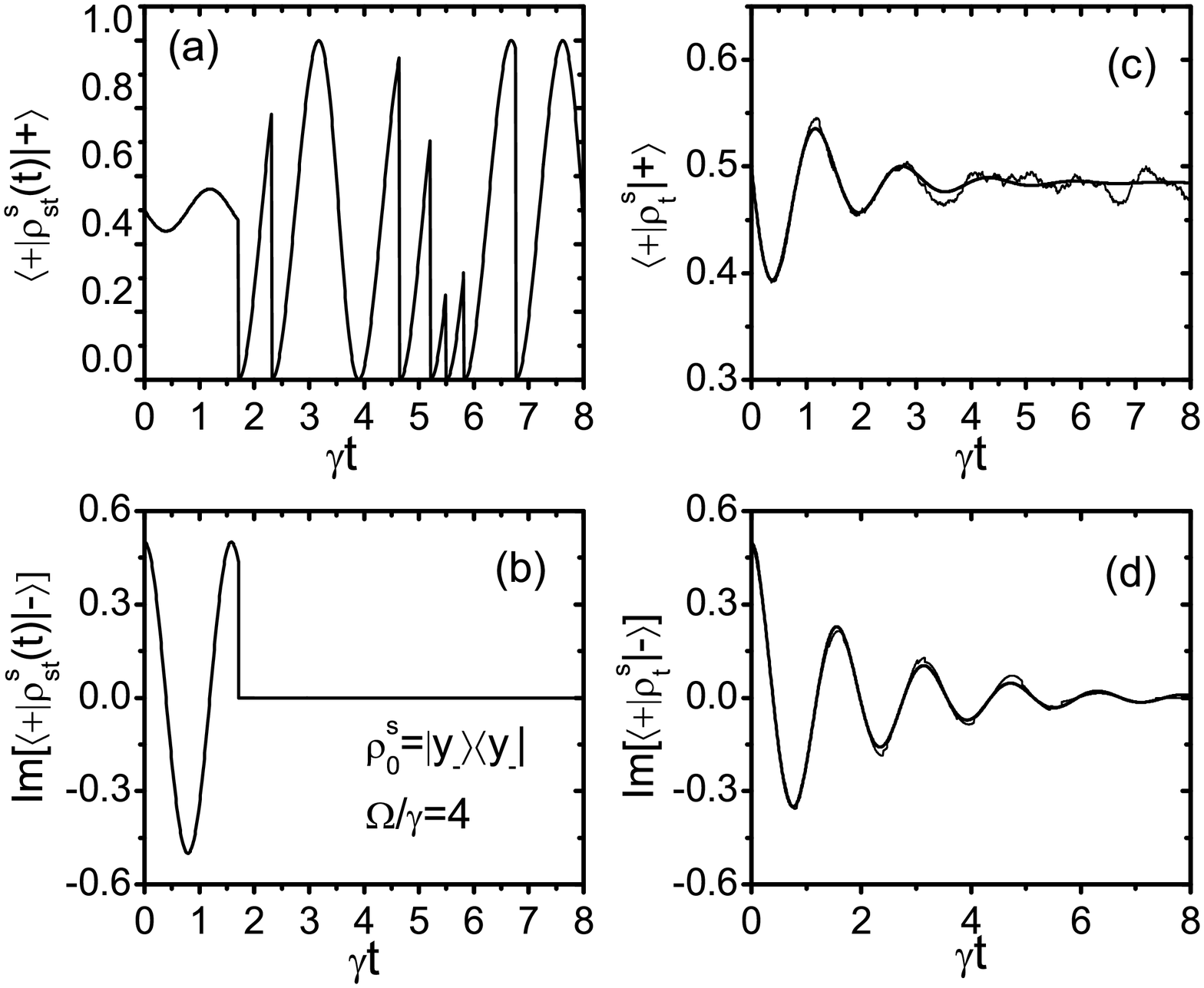}
\caption{Realizations of the stochastic density matrix $\protect\rho _{%
\mathrm{st}}^{s}(t)$\ and its ensemble average. In (a) and (b) are plotted
the population $\left\langle +\right\vert \protect\rho _{\mathrm{st}%
}^{s}(t)\left\vert +\right\rangle $\ and the imaginary part of the coherence 
$\left\langle +\right\vert \protect\rho _{\mathrm{st}}^{s}(t)\left\vert
-\right\rangle $ respectively. In (c) and (d) are plotted an average over $%
2\times 10^{3}$ realizations (noisy curves). The full lines correspond to
the analytical solutions Eqs. (\protect\ref{PuperSolution}) and (\protect\ref%
{CoherSolution}). In all cases the initial system state is $\protect\rho %
_{0}^{s}=\left\vert y_{-}\right\rangle \left\langle y_{-}\right\vert ,$
while the characteristic parameters satisfy $\Omega /\protect\gamma =4.$}
\end{figure}

In Figs. 2(c) and (d) we plot the population and coherence behaviors
obtained by averaging $2\times 10^{3}$ realizations (noisy curves). In
addition we also show the curves corresponding to the exact solution of the
density matrix evolution. For the chosen parameter values [$\gamma =\gamma
^{\prime }$ in Eq. (\ref{LindbladTLS})] it acquires the structure defined by
Eq. (\ref{LocalNonLocal}). By introducing the matrix elements%
\begin{equation}
p_{t}^{\pm }\equiv \left\langle \pm \right\vert \rho _{t}^{s}\left\vert \pm
\right\rangle ,\ \ \ \ \ c_{t}^{\pm }\equiv \left\langle \pm \right\vert
\rho _{t}^{s}\left\vert \mp \right\rangle ,  \label{MatrixElementsRhoS}
\end{equation}%
the evolution of the population can be written as in Eq. (\ref%
{PoblacionCondicional}) under the replacement $\tilde{p}_{t}^{\pm
}\rightarrow p_{t}^{\pm }$ and taking $\tilde{\delta}=0.$ Therefore, the
populations are governed by a memory-like classical rate equation. The
solution of these time convoluted evolutions read%
\begin{equation}
p_{t}^{+}\!=\!\frac{\Omega ^{2}}{\gamma ^{2}+2\Omega ^{2}}\!\Big{\{}\!1+e^{-%
\frac{3\gamma t}{4}}\!\Big{[}\mathrm{q}_{c}\cosh (\mu t)-\mathrm{q}_{s}\frac{%
\gamma }{\mu }\sinh (\mu t)\Big{]}\!\Big{\}},  \label{PuperSolution}
\end{equation}%
where%
\begin{equation}
\mu =\sqrt{(\gamma /4)^{2}-\Omega ^{2}}.
\end{equation}%
The dimensionless coefficient $\mathrm{q}_{c}$ and $\mathrm{q}_{s}$
introduce the dependence on the initial conditions 
\begin{subequations}
\begin{eqnarray}
\mathrm{q}_{c} &=&p_{0}^{+}(\gamma ^{2}/\Omega ^{2})+(p_{0}^{+}-p_{0}^{-}),
\\
\mathrm{q}_{s} &=&[p_{0}^{+}(\gamma ^{2}/\Omega
^{2})+(5p_{0}^{+}+3p_{0}^{-})]/4.
\end{eqnarray}%
The lower population follows as $p_{t}^{-}=1-p_{t}^{+}.$ The evolution of
the coherences can be written as in Eq. (\ref{CoherenciaCondicional}) after
replacing $\tilde{c}_{t}^{\pm }\rightarrow c_{t}^{\pm }.$ Their explicit
solution is 
\end{subequations}
\begin{equation}
c_{t}^{+}=e^{-\frac{\gamma t}{2}}\frac{1}{2}\left[ a-b\cosh (\nu t)\right] ,
\label{CoherSolution}
\end{equation}%
where the coefficients $a$ and $b$ read 
\begin{subequations}
\begin{eqnarray}
a &=&[c_{0}^{+}\gamma ^{2}/2-(c_{0}^{+}+c_{0}^{-})\Omega ^{2}]/\nu ^{2}, \\
b &=&(c_{0}^{+}-c_{0}^{-})\Omega ^{2}/\nu ^{2}.
\end{eqnarray}

In Figs. 2(c) and (d), the analytical expressions of both the populations
and coherences, Eqs. (\ref{PuperSolution}) and (\ref{CoherSolution})
respectively, recover the ensemble average behavior. This result explicitly
show the consistence of the proposed approach. On the other hand, Eqs. (\ref%
{PuperSolution}) and (\ref{CoherSolution}) lead to a diagonal stationary
density matrix 
\end{subequations}
\begin{equation}
\rho _{\infty }^{s}=\lim_{t\rightarrow \infty }\rho _{t}^{s}=\mathrm{diag}%
\Big{\{}\frac{\Omega ^{2}}{\gamma ^{2}+2\Omega ^{2}},\frac{\gamma
^{2}+\Omega ^{2}}{\gamma ^{2}+2\Omega ^{2}}\Big{\}}.
\end{equation}

The evolution of the matrix elements (\ref{MatrixElementsRhoS}) can also be
rewritten in terms of the system density matrix $\rho _{t}^{s}.$ From Eqs. (%
\ref{PoblacionCondicional}) ($\tilde{\delta}\rightarrow 0,$ $\tilde{p}%
_{t}^{\pm }\rightarrow p_{t}^{\pm },$ $\tilde{c}_{t}^{\pm }\rightarrow
c_{t}^{\pm }$) and (\ref{CoherenciaCondicional}) we find%
\begin{equation}
\frac{d\rho _{t}^{s}}{dt}=\gamma \mathcal{\hat{C}}[\sigma ]\rho
_{t}^{s}+\sum_{i=x,y,z}\int_{0}^{t}dt^{\prime }k_{t-t^{\prime }}^{i}\mathcal{%
\hat{C}}[\sigma _{i}]\rho _{t^{\prime }}^{s},  \label{LindbladMemorioso}
\end{equation}%
where the Lindblad channels are defined by Eq. (\ref{LindbladDEF}), $\sigma
_{i},$ $i=x,y,z,$ are the Pauli matrixes, while the memory functions are%
\begin{equation*}
k_{t}^{x}=\frac{\Omega ^{2}}{8}(e^{-\gamma t/2}+1)^{2},\ \ \ \ \
-k_{t}^{y}=k_{t}^{z}=\frac{\Omega ^{2}}{8}(e^{-\gamma t/2}-1)^{2}.
\end{equation*}%
As expected, the density matrix evolution (\ref{LindbladMemorioso}) has the
structure defined by Eq. (\ref{LocalNonLocal}), where the local in time
contribution is directly associated to the system transitions recorded by
the measurement apparatus.

\subsection*{Genuine non-Markovian effects}

Quantum non-Markovian time convoluted master equations can always be
rewritten in terms of local in time evolutions with time dependent rates 
\cite{Kosa}. If the rates are positive at all times, the measurement
dynamics is still consistent with a standard QJA \cite{maniscalco}. On the
other hand, if the rates assume negative values, the dynamics develops
\textquotedblleft genuine\textquotedblright\ non-Markovian effects such an
environment-to-system flow of information. This phenomenon can be detected
through different measures \cite{NoMeasure}, which in the Markovian case
present a monotonous time decay behavior \cite{breuerbook}. Now, we
demonstrate that this phenomenon can also arises in master equations such as
Eqs. (\ref{LocalNonLocal}) and (\ref{MasterRenewal}). As a measure we choose
the relative entropy \cite{breuerbook} with respect to the stationary state (%
$\rho _{s}^{\infty }=\lim_{t\rightarrow \infty }\rho _{t}^{s}$)%
\begin{equation}
E(\rho _{t}^{s}||\rho _{\infty }^{s})=\mathrm{Tr}_{s}[\rho _{t}^{s}(\ln
_{2}\rho _{t}^{s}-\ln _{2}\rho _{s}^{\infty })].  \label{Relativa}
\end{equation}%
%
%
%
\begin{figure}[tbp]
\includegraphics[bb=56 284 410 545,width=8.5cm]{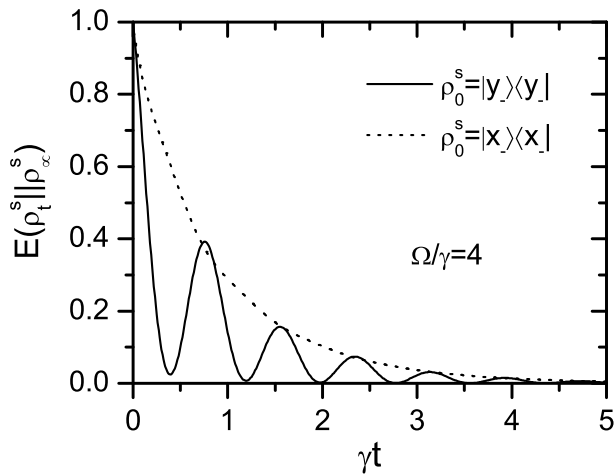}
\caption{Relative entropy with respect to the stationary state, Eq. (\protect
\ref{Relativa}). The density matrix follows from Eq. (\protect\ref%
{LindbladMemorioso}). For the full line the initial condition is $\protect%
\rho _{0}^{s}=\left\vert y_{-}\right\rangle \left\langle y_{-}\right\vert ,$
while for the dotted line is $\protect\rho _{0}^{s}=\left\vert
x_{-}\right\rangle \left\langle x_{-}\right\vert ,$ where $\left\vert
x_{-}\right\rangle =(1/\protect\sqrt{2})(\left\vert +\right\rangle
-\left\vert -\right\rangle ).$ In both cases $\Omega /\protect\gamma =4.$}
\end{figure}

In Fig. 3 the density matrix obey the evolution (\ref{LindbladMemorioso}),
whose solution is defined by Eqs. (\ref{PuperSolution}) and (\ref%
{CoherSolution}). The solid line corresponds to the initial condition and
parameters values of Figs. 1 and 2. Evidently, the oscillatory behavior of $%
E(\rho _{t}^{s}||\rho _{\infty }^{s})$ demonstrate that (\ref%
{LindbladMemorioso}) cannot be rewritten in terms of a local in time
evolution with (time-dependent) positive rates. The same property arises by
choosing the initial conditions $\rho _{0}^{s}=\left\vert \pm \right\rangle
\left\langle \pm \right\vert ,$ in which case the system dynamics can be
mapped with a classical two-level system. In general, the development of or
not of the revivals strongly depends on the initial conditions. For example,
for $\rho _{0}^{s}=\left\vert x_{-}\right\rangle \left\langle
x_{-}\right\vert ,$ where $\left\vert x_{-}\right\rangle $ is an eigenvector
of $\sigma _{x}$ with eigenvalue minus one, $\left\vert x_{-}\right\rangle
=(1/\sqrt{2})(\left\vert +\right\rangle -\left\vert -\right\rangle ),$ $%
E(\rho _{t}^{s}||\rho _{\infty }^{s})$ decay in a monotonous way (dotted
line). This case can be understood in terms of the symmetries of the
underlying bipartite dynamics, Eq. (\ref{LindbladTLS}).

\section{Summary and conclusions}

In this paper we established a non-Markovian generalization of the standard
QJA. The underlying idea consist in embedding the system dynamics in a
bipartite Markovian evolution [Eq. (\ref{LindbladBipartito})]. Assuming that
the measurement action is only performed on the system of interest, we
demonstrated that there exist symmetries conditions on the Lindblad
(bipartite) channels [Eqs. (\ref{rateCondition}) and (\ref%
{rateConditionNoRenewal})] that lead to a closed system stochastic dynamics
consistent with a quantum measurement theory.

For both, renewal and non-renewal measurement processes, the ensemble of
realizations is similar to that of the standard case. At random times, the
system state suffer a disruptive transformation, which is associated to each
recording event. In the intermediate time intervals, the (conditional)
system dynamic is smooth and non-unitary. The main difference with the
standard approach is this last ingredient. Here, it is not defined by an
exponential propagator [Eq. (\ref{d/dtT(t)})]. In fact, it arises from a
partial trace over the semigroup evolution associated to the Markovian
bipartite dynamics [Eq. (\ref{PropCondNoMarkov})]. Hence, in general, the
stochastic dynamics does not admit an unravelling in terms of pure states
[Eq. (\ref{NoSWF})].

As in the standard case, the jump statistics can be defined by a survival
probability [Eq. (\ref{SurvivalNoMarkov})], which in general depends on the
system state. In addition to the stochastic dynamics, we also characterized
the system density matrix evolution. The structure of the corresponding
non-Markovian quantum master equations is defined by Eqs. (\ref%
{LocalNonLocal}) and (\ref{MasterRenewal}). Arbitrary master equations with
this structure can be unravelled with the ensemble of trajectories if it is
possible to assign a survival probability to the conditional dynamics, Eq. (%
\ref{PropagatorConditionSur}).

The consistence of the formalism was checked by studying the dynamics of a
two level system whose non-Markovian dynamics lead to successive transition
between the upper and lower levels. The simplicity of the model allowed us
to obtain short analytical expressions for both the measurement statistics
[Eq. (\ref{SurvivalAnalitica})] as well as for the density matrix elements
and the corresponding density matrix evolution [Eq. (\ref{LindbladMemorioso}%
)]. The relevance of the example not only come from its simplicity. In fact,
it also allowed us to demonstrate that the present generalization is
consistent with a back flow of information from the environment to the
system. This property follows from the non-monotonous decay of the relative
entropy with respect to the stationary state (Fig. 3).

While the present formalism lead to a consistent non-Markovian
generalization of the quantum jumps approach, it is clear that it can be
extended in different directions. For example one may consider arbitrary
initial bipartite states [Eq. (\ref{Inicial})] or to introduce non-separable
bipartite resetting states [Eq. (\ref{Mapping})]. A less technical aspect
should be to consider the case in which many different measurement apparatus
are monitoring the system or to determine which kind of consistent
non-Markovian generalization is not covered by a Markovian embedding.

\section*{Acknowledgments}

This work was supported by CONICET, Argentina, under Grant No. PIP
11420090100211.

\appendix

\section{Quantum jumps statistics-Markovian case}

Here we derive the statistical description of the ensemble of realizations
associated to the Markovian QJA. The solution of Eq. (\ref{LindbladSA}) can
formally be written as%
\begin{equation}
\rho _{t}^{s}=\exp [\mathcal{\hat{D}}t]\rho _{0}^{s}+\int_{0}^{t}dt^{\prime
}\exp [\mathcal{\hat{D}}(t-t^{\prime })]\mathcal{\hat{J}}[\rho _{t^{\prime
}}^{s}],
\end{equation}%
where $\rho _{0}^{s}$\ is the initial system state. This expression can be
iterated leading to the series expansion%
\begin{equation}
\rho _{t}^{s}=\sum_{n=0}^{\infty }\rho _{t}^{(n)},  \label{Unravelling}
\end{equation}%
where each contribution satisfies the recursive relation%
\begin{equation}
\rho _{t}^{(n)}=\int_{0}^{t}dt^{\prime }\mathcal{\hat{T}}(t-t^{\prime })%
\mathcal{\hat{J}}\rho _{t^{\prime }}^{(n-1)},  \label{RecursivaMarkov}
\end{equation}%
with $\rho _{t}^{(0)}=\mathcal{\hat{T}}(t)\rho _{0}^{s}.$ Therefore, it
follows $(n\geq 1)$%
\begin{equation}
\rho _{t}^{(n)}\!=\!\int_{0}^{t}dt_{n}\cdots \!\int_{0}^{t_{2}}dt_{1}\ 
\mathcal{\hat{T}}(t-t_{n})\mathcal{\hat{J}}\cdots \mathcal{\hat{T}}%
(t_{2}-t_{1})\mathcal{\hat{J}\hat{T}}(t_{1})\rho _{0}^{s}.
\end{equation}%
The superoperators $\mathcal{\hat{J}}$ and $\mathcal{\hat{T}}(t)$ are
defined by Eqs. (\ref{JMarkov}) and (\ref{Texponencial}) respectively. Each
contribution $\rho _{t}^{(n)}$ can be associated to trajectories with $n$%
-detection events. Its statistics can be obtained by writing the previous
expression in terms of the measurement transformation $\mathcal{\hat{M}}$
[Eq. (\ref{MOperator})] and the normalized propagator $\mathcal{\hat{T}}%
_{c}(t)$ [Eq. (\ref{Conditional})]. We get%
\begin{eqnarray}
\rho _{t}^{(n)}\! &=&\!\int_{0}^{t}dt_{n}\cdots \!\int_{0}^{t_{2}}dt_{1}\
P_{n}[t,\{t_{i}\}_{1}^{n}]  \label{RhoEneEstadistica} \\
&&\times \mathcal{\hat{T}}_{c}(t-t_{n})\mathcal{\hat{M}}\cdots \mathcal{\hat{%
T}}_{c}(t_{2}-t_{1})\mathcal{\hat{M}\hat{T}}_{c}(t_{1})\rho _{0}^{s},\ \ \ \
\ \ \   \notag
\end{eqnarray}%
$(n\geq 1)$ and $\rho _{t}^{(0)}=P_{0}(t|\rho _{0}^{s})\mathcal{\hat{T}}%
_{c}(t)\rho _{0}^{s}.$ The function%
\begin{equation}
P_{n}[t,\{t_{i}\}_{1}^{n}]=\mathrm{Tr}_{s}[\mathcal{\hat{T}}(t-t_{n})%
\mathcal{\hat{J}}\cdots \mathcal{\hat{J}\hat{T}}(t_{2}-t_{1})\mathcal{\hat{J}%
\hat{T}}(t_{1})\rho _{0}^{s}],  \label{ConjuntaCruda}
\end{equation}%
is the joint probability density for observing measurement events at times $%
\{t_{i}\}_{1}^{n}.$ It completely characterize the statistic of the
measurement process. By introducing the auxiliary states $\rho _{t_{i+1}}=%
\mathcal{\hat{T}}_{c}(t_{i+1},t_{i})\mathcal{\hat{M}}\rho _{t_{i}},$ with $%
\rho _{t_{1}}=\mathcal{\hat{T}}_{c}(t_{1},0)\rho _{0},$ the previous object
can be rewritten as%
\begin{eqnarray}
P_{n}[t,\{t_{i}\}_{1}^{n}] &=&P_{0}(t-t_{n}|\mathcal{\hat{M}}\rho _{t_{n}})
\label{Joint} \\
&&\prod_{j=2}^{n}w(t_{j}-t_{j-1}|\mathcal{\hat{M}}\rho
_{t_{j-1}})w(t_{1}|\rho _{0}^{s}),  \notag
\end{eqnarray}%
where the survival probability $P_{0}(t|\rho )$ and the waiting time
distribution $w(t|\rho )$ are defined by Eqs. (\ref{SurMarkov}) and (\ref%
{WaitMarkov}) respectively.

The structure of both Eqs. (\ref{RhoEneEstadistica}) and (\ref{Joint}) are
consistent with the stochastic dynamics defined in Sec. II-A. The second
line of Eq. (\ref{RhoEneEstadistica}) consists in successive applications of
the measurement transformations $\mathcal{\hat{M}}$ and intermediate
evolution with the propagator $\mathcal{\hat{T}}_{c}(t).$ On the other hand,
the weight of each realization, defined by Eq. (\ref{Joint}), have the same
structure than a renewal process, that is, there exist a probability
distribution (waiting time distribution) that define the statistic of the
time interval between consecutive detection events. Nevertheless, here the
distribution depends on the resetting state, that is, the state after a
measurement event.

\section{Non-Markovian master equations from the jumps statistics}

We derived the non-Markovian extension of the QJA by studying the standard
approach in a Markovian bipartite dynamics. Under the conditions obtained in
Sec. III the system stochastic dynamics becomes closed, that is, it can be
written without taking into account explicitly the ancilla dynamics. Here we
derive the corresponding non-Markovian master equation [see Eqs. (\ref%
{LocalNonLocal}) and (\ref{MasterRenewal})] by averaging the ensemble of
trajectories.

The full counting statistics can be derived from the Markovian evolution Eq.
(\ref{BipartitaD+JMarkkov}) and its formal solution (\ref{RhoSAIntegrada}).
All calculation steps described in Appendix A can be extended, after a
trivial change of notation ($\mathcal{\hat{J}}\rightarrow \mathbb{\hat{J}},\ 
\mathcal{\hat{T}}\rightarrow \mathbb{\hat{T}}$), to the bipartite evolution
defined in terms of $\rho _{t}^{sa}.$ By performing a partial trace over the
ancilla degrees of freedom on the corresponding expressions, by using the
bipartite measurement transformation (\ref{Mapping}) and the initial
bipartite state (\ref{Inicial}), it is possible to demonstrate that Eqs. (%
\ref{RhoEneEstadistica}) and (\ref{Joint}) are also valid for the
non-Markovian system dynamics. Nevertheless, in the non-Markovian case, the
propagator $\mathcal{\hat{T}}(t)$ is defined by Eq. (\ref{PropCondNoMarkov})
[or equivalently Eq. (\ref{d/dtT(t)})] while the survival probability $%
P_{0}(t|\rho )$\ and waiting time distribution $w(t|\rho )$ from Eqs. (\ref%
{SurvivalNoMarkov}) and (\ref{waitingNoMarkov}) respectively.

\subsection{Renewal case}

When the measurement process is a renewal one, we can write the joint
probability density [Eq. (\ref{Joint})] as%
\begin{equation}
P_{n}[t,\{t_{i}\}_{1}^{n}]=P_{0}(t-t_{n})%
\prod_{j=2}^{n}w(t_{j}-t_{j-1})w(t_{1}|\rho _{0}^{s}),
\end{equation}%
where, in contrast to a Markovian renewal process, here the survival
probability\ $P_{0}(t)$\ and waiting time distribution $w(t)$ are defined by
Eqs. (\ref{SurNoMarkRenewal}) and (\ref{WaiterNoMarkRenewal}) respectively.
From Eq. (\ref{RhoEneEstadistica}) and by using the renewal property Eq. (%
\ref{Mrenewal}), the previous expression for $P_{n}[t,\{t_{i}\}_{1}^{n}]$
allows us to write%
\begin{equation}
\rho _{t}^{(n)}=\int_{0}^{t}dt^{\prime }\mathcal{\hat{T}}(t-t^{\prime })\bar{%
\rho}_{s}\ f^{(n)}(t^{\prime }),  \label{Rho(n)NoMark}
\end{equation}%
$[\rho _{t}^{(0)}=\mathcal{\hat{T}}(t)\rho _{0}^{s}]$ where the function $%
f^{(n)}(t)$\ is defined as%
\begin{equation}
f^{(n)}(t)=\int_{0}^{t}dt_{n}\cdots
\int_{0}^{t_{2}}dt_{1}\prod_{j=2}^{n}w(t_{j}-t_{j-1})w(t_{1}|\rho _{0}^{s}).
\end{equation}%
From Eq. (\ref{Rho(n)NoMark}) and the expression for the waiting time
distribution $w(t)$ [Eq. (\ref{WaiterNoMarkRenewal})], we get the recursive
relation%
\begin{equation}
\rho _{t}^{(n)}=-\int_{0}^{t}dt^{\prime }\mathcal{\hat{T}}(t-t^{\prime })%
\bar{\rho}_{s}\int_{0}^{t^{\prime }}dt^{\prime \prime }\mathrm{Tr}_{s}[%
\mathcal{\hat{D}}(t^{\prime }-t^{\prime \prime })\rho _{t^{\prime \prime
}}^{(n-1)}].
\end{equation}%
By adding all these states [see Eq. (\ref{Unravelling})] and by using the
non-Markovian time evolution of the propagator $\mathcal{\hat{T}}(t)$\ [Eq. (%
\ref{d/dtT(t)})], after some calculations steps, the system density matrix
evolution Eq. (\ref{MasterRenewal}) is recovered.

\subsection{Non-renewal case}

By using the rate condition Eq. (\ref{rateConditionNoRenewal}) correspondent
to the non-renewal case, it is possible to demonstrate that the
superoperator $\mathbb{\hat{J}}$ [Eq. (\ref{JBipartito})] satisfies the
relation%
\begin{equation}
\mathbb{\hat{J}}\rho =\mathcal{\hat{J}}\{\mathrm{Tr}_{a}[\rho ]\}\otimes 
\bar{\rho}_{a},  \label{JJ}
\end{equation}%
where the system superoperator $\mathcal{\hat{J}}$ is defined by Eq. (\ref%
{JMarkov}) and the ancilla resetting state $\bar{\rho}_{a}$ follows from Eq.
(\ref{AncillaReseting}).

By writing Eq. (\ref{ConjuntaCruda}) in terms of bipartite objects ($%
\mathcal{\hat{J}}\rightarrow \mathbb{\hat{J}},\ \mathcal{\hat{T}}\rightarrow 
\mathbb{\hat{T}}$), after introducing Eq. (\ref{JJ}), the joint probability
distribution can be written as%
\begin{equation}
P_{n}[t,\{t_{i}\}_{1}^{n}]=\mathrm{Tr}_{s}[\mathcal{\hat{T}}(t-t_{n})%
\mathcal{\hat{J}}\cdots \mathcal{\hat{J}\hat{T}}(t_{2}-t_{1})\mathcal{\hat{J}%
\hat{T}}(t_{1})\rho _{0}^{s}],  \label{JointNoRenewal}
\end{equation}%
where $\mathcal{\hat{T}}(t)$ and $\mathcal{\hat{J}}$\ follows from Eqs. (\ref%
{PropCondNoMarkov}) and (\ref{JMarkov}) respectively. Notice that in this
case, the only difference with the Markovian case [Eq. (\ref{ConjuntaCruda}%
)] is the definition of $\mathcal{\hat{T}}(t).$

In order to obtain the density matrix evolution we need a recursive relation
for the states $\rho _{t}^{(n)}.$ Here, such kind of relation can be easily
obtained from the recursive relation (\ref{RecursivaMarkov}) when applied to
the bipartite dynamics. With the aid of Eq. (\ref{JJ}) we get%
\begin{equation}
\rho _{t}^{(n)}=\int_{0}^{t}dt^{\prime }\mathcal{\hat{T}}(t-t^{\prime })%
\mathcal{\hat{J}}\rho _{t^{\prime }}^{(n-1)}.
\end{equation}%
Consistently, the same relation arise from Eqs. (\ref{RhoEneEstadistica})
and (\ref{JointNoRenewal}). By adding all states $\rho _{t}^{(n)},$ and by
using the non-Markovian time evolution of the propagator $\mathcal{\hat{T}}%
(t)$\ [Eq. (\ref{d/dtT(t)})], we recover Eq. (\ref{LocalNonLocal}).

\end{document}